\documentclass[journal,onecolumn]{IEEEtran}

\usepackage{amssymb,bm,amsthm,color,amsmath,url,multirow}
\usepackage{cite}
\newtheorem*{claim}{Claim}
\newtheorem{thm}{Theorem}[section]
\newtheorem{lem}{Lemma}[section]
\newtheorem{dfn}{Definition}[section]
\newtheorem{prop}{Proposition}[section]
\newtheorem{rmk}{Remark}[section]
\newtheorem{cor}{Corollary}[section]
\newtheorem{example}{Example}[section]
\newtheorem{clm}{Claim}
\newenvironment{pf}{{\noindent\it Proof:}}{\hfill $\blacksquare$\par}

\newcommand{\RNum}[1]{\lowercase\expandafter{\romannumeral #1\relax}}
\newcommand{\Rnum}[1]{\uppercase\expandafter{\romannumeral #1\relax}}

\newcommand{\tabincell}[2]{\begin{tabular}{@{}#1@{}}#2\end{tabular}}

\begin{document}

\title{Some New Results on Splitter Sets}

\author{Zuo Ye, Tao Zhang, Xiande Zhang and Gennian Ge
\thanks{The research of Z. Ye and X. Zhang were supported by the National Natural Science Foundation of China under grant 11771419.}
\thanks{The research of T. Zhang was supported by the National Natural Science Foundation of China under grant 11801109.}
\thanks{The research of G. Ge was supported by the National Natural Science Foundation of China under Grant Nos. 11431003, 61571310 and 11971325, Beijing Scholars Program, Beijing Hundreds of Leading Talents Training Project of Science and Technology, and Beijing Municipal Natural Science Foundation.}
\thanks{Z. Ye and X. Zhang are with  the School of Mathematical Sciences, University of Science and Technology of China, Hefei 230026, Anhui, China (e-mails: zyprince@mail.ustc.edu.cn; drzhangx@ustc.edu.cn).}
\thanks{T. Zhang is with the School of Mathematics and Information Science, Guangzhou University, Guangzhou 510006, China (e-mail: taozhang@gzhu.edu.cn).}
\thanks{G. Ge is with  the School of Mathematical Sciences, Capital Normal University,
Beijing 100048, China (e-mail: gnge@zju.edu.cn).}
}

\maketitle
\begin{abstract}
Splitter sets have been widely studied due to their applications in flash memories, and their close relations with lattice tilings and conflict avoiding codes. In this paper, we give necessary and sufficient conditions for the existence of nonsingular perfect splitter sets, $B[-k_1,k_2](p)$ sets,  where $0\le k_{1}\leq k_{2}=4$. Meanwhile, constructions of nonsingular perfect splitter sets are given. When perfect splitter sets do not exist, we present four new constructions of quasi-perfect splitter sets. Finally, we give a connection between nonsingular splitter sets and Cayley graphs, and as a byproduct, a general lower bound on the maximum size of nonsingular splitter sets is given.
\end{abstract}

\section{Introduction}
Flash memory is a non-volatile, high-density and low-cost memory. There are many fields in which flash memory has found its applications, such as personal computers, digital audio players, digital cameras, mobile phones, embedded systems and so on.
\par
A multilevel flash cell is electrically programmed into one of $q$ threshold states and therefore can be regarded as storing one symbol from the set $\mathbb{Z}_q$. Many reported common flash error mechanisms induce errors whose magnitudes are small and  independent of the alphabet size, which may be significantly larger than the typical error magnitude. Thus, flash errors gave a strong motivation for the application of the limited magnitude error model to flash memory \cite{wz7,wz8}.
\par
Splitter sets were first studied in \cite{wz9,wz10,wz11,bk4,wz12,wz13,wz14} in the language of lattice tilings. Recently, an important application to the limited magnitude error-correcting codes for flash memories has been found \cite{wz7,wz8}. In this context, a code obtained from a splitter set $B[-k_1,k_2](q)$ can correct a symbol $a\in\{0,1,\cdots,q-1\}$ if it is modified into $(a+e)\pmod{q}$ during transmission, where $-k_1\le e\le k_2$. This new finding has immediately motivated a lot of research on splitter sets (see \cite{wz15,wz16,wz7,wz5,wz17,wz18,wz19,wz20,wz2,wz1} and the references therein). Moreover, splitter sets are also useful in the constructions of conflict avoiding codes and $k$-radius sequences \cite{wz3,wz21}.
\par
Some researchers considered the existence of perfect splitter sets. In \cite{wz5}, the authors presented a construction of perfect splitter sets for $k_1=0$. The existence of perfect splitter sets for $k_1=k_2$ has been studied in \cite{wz17,bk4}. Some constructions of perfect splitter sets for $1\le k_1<k_2$ were given in \cite{wz19,wz2,wz3}. For the nonexistence results, Woldar \cite{wz22} obtained some necessary conditions for the existence of purely singular perfect splitter sets for $k_1=0$. In \cite{wz19,wz20}, Schwartz gave some necessary conditions for the existence of perfect splitter sets for more general $1\le k_1<k_2$. In \cite{wz23,wz1,wz3}, the authors proved that there does not exist a nonsingular perfect splitter set when $1\le k_1<k_2$ and $k_1+k_2$ is an odd integer. In \cite{wz4}, Yuan and Zhao gave a necessary and sufficient condition for the existence of nonsingular perfect $B[-1,3](p)$ sets. For $k_{1}=k_{2}=4$, Tamm \cite{wz14} provided a list of primes $p$ for which a perfect $B[-4,4](p)$ set exists. In \cite{M95}, Munemasa pointed out that the subgroup $\langle-1,2,3\rangle$ plays a central role in the study of perfect $B[-4,4](p)$ sets.
\par
Since perfect splitter sets only exist for certain parameters, other researchers also studied quasi-perfect splitter sets and optimal splitter sets. In \cite{wz7}, Kl\o ve et al. gave a construction of quasi-perfect splitter sets for $k_1=0$. Some constructions of quasi-perfect splitter sets for $k_1=k_2$ can be found in \cite{wz17}. The authors of \cite{wz1} gave a construction of quasi-perfect splitter sets for $1\le k_1<k_2$. The exact size of maximal $B[-k_1,k_2](q)$ sets for $0\le k_{1}\le k_{2}\le4$ and certain $q$  can be found in \cite{wz7,wz17,wz2,wz24,wz25}.
\par
In this work, we continue to derive new results for splitter sets. We give necessary and sufficient conditions for the existence of nonsingular perfect $B[-k_{1},k_{2}](p)$ sets, where $(k_{1},k_{2})\in\{(0,4),(2,4),(4,4)\}$. We also present four new constructions of quasi-perfect splitter sets.
This paper is organized as follows. In Section \ref{sec2}, we introduce some notations and terminologies which will be used throughout the paper. In Section \ref{sec3}, we give some necessary and sufficient conditions for the existence of nonsingular perfect $B[-k_{1},k_{2}](p)$ sets, where $(k_{1},k_{2})\in\{(0,4),(2,4),(4,4)\}$. In Section \ref{sec4}, four new constructions of quasi-perfect splitter sets are presented. In Section \ref{sec5}, we give a connection between nonsingular splitter sets and Cayley graphs, and by product, a lower bound on the maximum size of nonsingular splitter sets is given. Finally, section \ref{sec6} concludes the paper.

\section{Preliminary}\label{sec2}
In this section, we introduce some useful notations and terminologies, and recall several relevant results which will be used later.

 For integers $m,n$ such that $m\le n$, we denote $[m,n]:=\{m,m+1,\ldots,n\}$ and $
[m,n]^{*}:=\{m,m+1,\ldots,n\}\setminus\{0\}$. For any integer $q\ge 2$, let $\mathbb{Z}_q$ be the ring of integers modulo $q$ and let $\mathbb{Z}$ be the ring of integers. If $a$ is an element of $\mathbb{Z}_q$ and $S$ is a subset of integers, then $aS$ denotes the set $\{as \pmod q :s\in S\}$.

\begin{dfn}
Let $q,k_1,k_2\in\mathbb{Z}$ with $q\ge 2$ and $0\le k_1\le k_2$. The set $B\subset\mathbb{Z}_q$ is called a \emph{splitter set} if each of the sets
$b\left[-k_{1}, k_{2}\right]^{*}$, $b \in B$,
has $k_1+k_2$ nonzero elements, and they are pairwise disjoint. We denote such a splitter set by a $B[-k_1,k_2](q)$ set.
\end{dfn}
%\begin{align*}
%\left\{a b \pmod{q}: \text{ }a \in\left[-k_{1}, k_{2}\right]^{*}\right\},\text{ } b \in B
%\end{align*}

From the definition, if $B$ is a $B[-k_1,k_2](q)$ set, then $|B|\le\frac{q-1}{k_1+k_2}$. If $|B|=\frac{q-1}{k_1+k_2}$, then we say that $B$ is \emph{perfect}. It is clear that a perfect $B[-k_1,k_2](q)$ set exists only if $q\equiv 1\pmod{k_1+k_2}$. If $q\not\equiv 1\pmod{k_1+k_2}$ and $|B|=\left\lfloor\frac{q-1}{k_1+k_2}\right\rfloor$, then we say $B$ is \emph{quasi-perfect}.
A perfect $B[-k_1,k_2](q)$ set is called \emph{nonsingular} if $\gcd(q,k_2!)=1$. Otherwise, it is called \emph{singular}. The following theorems can be found in \cite{wz20,wz2}.
\begin{thm}\textup{\cite[Theorem 14]{wz20}}
Suppose that  there exists a perfect $B[-k_1,k_2](q)$ set. Then for any positive integer $d\mid q$ satisfying $\gcd(d,k_2!)=1$,  there is a perfect $B[-k_1,k_2](\frac{q}{d})$ set.
\end{thm}
\begin{thm}\textup{\cite[Theorem 5]{wz2}}
Let $B_{1}$ be a $B\left[-k_{1}, k_{2}\right]\left(q_{1}\right)$ set and $B_{2}$ be a $ B\left[-k_{1}, k_{2}\right]\left(q_{2}\right)$ set, where $\gcd\left(q_{2}, k_{2} !\right)=1$. Let
$$
B_{1} \odot B_{2}=\left\{c+r q_{1}:\text{ } c \in B_{1}, r \in\left[0, q_{2}-1\right]\right\} \cup\left\{q_{1} c: \text{ } c \in B_{2}\right\}.
$$
Then
\begin{enumerate}
  \item $B_{1} \odot B_{2}$ is a $B\left[-k_{1}, k_{2}\right]\left(q_{1} q_{2}\right)$ set;
  \item $\left|B_{1} \odot B_{2}\right|=q_{2}\left|B_{1}\right|+\left|B_{2}\right|$;
  \item If both $B_{1}$ and $B_{2}$ are perfect, then $B_{1} \odot B_{2}$ is perfect.
\end{enumerate}
\end{thm}
From the above two theorems, it is easy to see that there is a perfect nonsingular $B[-k_1,k_2](q)$ set if and only if there
is a perfect nonsingular $B[-k_1,k_2](p)$ set for each prime factor $p$ of $q$. Therefore, in Section \ref{sec3}, when we deal with the existence of nonsingular perfect $B[-k_1,k_2](p)$ sets, we only consider the case when $p$ is a prime. In this case,
$\mathbb{Z}_p^{*}:=\mathbb{Z}_p\setminus \{0\}$ is a cyclic multiplicative group, so  we don't distinguish between integers and ring elements. The following necessary condition for the existence of perfect splitter sets is quite useful, which will be used frequently later.
\begin{lem}\textup{\cite[Lemma 2.4]{wz1}}\label{uniqint}
Let $k_1,k_2\in\mathbb{Z}$ with $0\le k_1\le k_2$, and $p$ be a prime. If $B$ is a perfect $B[-k_1,k_2](p)$ set, then for any $a\in\mathbb{Z}_p^{*}$, we have $|B\cap a[-k_1,k_2]^{*}|=1$.
\end{lem}
The definition of splitter sets is closely related to the following definition from group theory.
\begin{dfn}
  Let $(G,\cdot)$ be a finite group and $A,B$ be subsets of $G$. If for any element $g\in G$, there are unique elements $a\in A$ and $b\in B$ such that $g=a\cdot b$, then we say $G=A\cdot B$ is a \emph{factorization} of $G$, and $A$ (or $B$) is a \emph{direct factor} of $G$.
\end{dfn}
\begin{rmk}
When $p>k_1+k_2$ is a prime, we can view $[-k_1,k_2]^{*}$ as a subset of $\mathbb{Z}_p^{*}$. Then by the definition of perfect splitter sets and factorization, we see that $B$ is a perfect $B[-k_1,k_2](p)$ set if and only if $\mathbb{Z}_p^{*}=B[-k_1,k_2]^{*}$ is a factorization.
\end{rmk}

For convenience, we introduce more notations before closing this section.
 For a group $G$ and a subset $S\subseteq G$, $\langle S\rangle$ denotes the subgroup of $G$ generated by $S$. Suppose $g$ is a generator of $\mathbb{Z}_p^{*}$, then we say that $g$ is a {\it primitive root} modulo $p$. For any element $b\in \mathbb{Z}_p^{*}$, there exists a unique integer $i\in[0,p-2]$ such that $g^{i}\equiv b\pmod{p}$. We say $i$ is the {\it index of $b$ relative to the base $g$}, and denote it by $\mathrm{ind}_g(b)$. If $x\in\mathbb{Z}_p^{*}$, let $\mathrm{ord}_p(x)$ denote the order of $x$ modulo $p$.

\section{Nonsingular perfect $B[-k_1,k_2](p)$ sets for $k_2= 4$}\label{sec3}
This section serves to provide complete characterizations of the existence of nonsingular perfect $B[-k_1,k_2](p)$ sets, where $(k_{1},k_{2})\in\{(0,4),(2,4),(4,4)\}$, and $p\equiv 1\pmod{k_1+k_2}$ is a prime. Since there does not exist a nonsingular perfect splitter set when $1\le k_1<k_2$ and $k_1+k_2$ is an odd integer by \cite{wz23}, we thus  completely solve the case when $k_2=4$.
%\begin{itemize}
%\item For any integers $a_1,a_2,\ldots,a_n$, where $n\ge 2$, let $\gcd(a_1,a_2,\ldots,a_n)$ be the greatest common divisor of $a_1,a_2,\ldots,a_n$.
%
%\end{itemize}
\subsection{Complete characterizations}
\begin{lem}\label{yl1}
Let $k\ge 1$ be an integer, $p\equiv 1\pmod{2k+2}$ be a prime,  and let $B$ be a nonsingular perfect $B[-k,k+2](p)$ set. If $i\in B$, then
$$
i\left<-\frac{k+1}{k+2}\right>\subseteq B,
$$
where $\left<-\frac{k+1}{k+2}\right>$ denotes the subgroup of $\mathbb{Z}_p^{*}$ generated by $-\frac{k+1}{k+2}$. In particular, the order of $-\frac{k+1}{k+2}$ in $\mathbb{Z}_p^{*}$ is odd.
\end{lem}
\begin{pf}
Since $B$ is a nonsingular perfect $B[-k,k+2](p)$ set, we have that for any $a\in\mathbb{Z}_p^{*}$, $|B\cap a[-k,k+2]^{*}|=1$ by Lemma~\ref{uniqint}. For any $i\in B$, taking $a=i$, we have
 \begin{align}\label{eqc}
 \left|B\cap i[-k,k+2]^{*}\right|=1.
 \end{align}
Since $i\in B\cap i[-k,k+2]^{*}$ and $-i\in i[-k,k+2]^{*}$, we get $-i\notin B$.
 Further taking $a=\pm\frac{i}{k+2}$, we have
 \begin{align}
&\left|B\bigcap \frac{i}{k+2}[-k,k+2]^{*}\right|=1, \label{eqa}\\
 &\left|B\bigcap \left(-\frac{i}{k+2}\right)[-k,k+2]^{*}\right|=1.\label{eqb}
 \end{align}
By (\ref{eqa}) and the fact that $i\in B\bigcap  \frac{i}{k+2}[-k,k+2]^{*}$, we get $B\bigcap \frac{i}{k+2}[-k,k+1]^{*}=\emptyset$. Observing that $ \left(-\frac{i}{k+2}\right)[-k,k+2]^{*}=\left (\frac{i}{k+2}[-k,k]^{*}\right)\bigcup \{-\frac{k+1}{k+2}i,-i\}$,  we have $-\frac{k+1}{k+2}i\in B$ by (\ref{eqb}). Then  replacing $i$ in (\ref{eqc}), (\ref{eqa}) and (\ref{eqb}) by $-\frac{k+1}{k+2}i$, and following the same arguments, we can get $i\left(-\frac{k+1}{k+2}\right)^2\in B$. Repeating this  procedure, we deduce that $i\left<-\frac{k+1}{k+2}\right>\subseteq B$. If $\mathrm{ord}_p(-\frac{k+1}{k+2})$ is even, then $-1\in\left<-\frac{k+1}{k+2}\right>$, hence $-i\in B$, which is a contradiction.
\end{pf}

The next lemma can be derived from Lemmas 2.3 and 2.5 of \cite{bk1}. We sketch the proof here to explain how to get a perfect splitter set.
\begin{lem}\label{yl2}
  Let $ k_2\geq k_1\geq 0$ be integers, $p\equiv 1\pmod{k_1+k_2}$ be a prime, and  $M=[-k_1,k_2]^{*}$. Then there is a nonsingular perfect $B[-k_1,k_2](p)$ set if and only if $M$ is a direct factor of the subgroup $H=\left<-1,2,\ldots,k_2\right>\subset \mathbb{Z}_p^{*}$.
\end{lem}
\begin{pf}
Suppose $B$ is a nonsingular perfect $B[-k_1,k_2](p)$ set. Let $B^{\prime}=B\cap H$, then it is easy to verify that $H=MB^{\prime}$ is a factorization.
\par
Now  suppose $H=MB^{\prime}$ is a factorization. Let $\{b_1,\cdots,b_s\}$ be a complete set of coset representatives of $H$ in $\mathbb{Z}_p^{*}$. Then $B=\mathop{\cup}\limits_{i=1}^{s}b_iB^{\prime}$ is a perfect $B[-k_1,k_2](p)$ set.
\end{pf}
\par
~\\
\indent
We are now ready to present our main results.
\begin{thm}\label{thm0}
  Let $p\equiv 1\pmod{6}$ be a prime. Then there is a nonsingular perfect $B[-2,4](p)$ set if and only if $\mathrm{ord}_p(-\frac{3}{4})$ is odd and $2\notin\langle 6,8\rangle$.
\end{thm}
\begin{pf}
 The necessity is just a combination of Lemma \ref{yl1} and Theorem 5.8 of \cite{wz1}. Now we consider the other direction. It is easy to check that
$$
(-1)^x2^y3^z=\left\{
\begin{array}{ll}
  (-1)^{x+\frac{y-z}{3}}\cdot 6^{\frac{y+2z}{3}}\left(-\frac{4}{3}\right)^{\frac{y-z}{3}}, & \text{if }y\equiv z\pmod{3},\\
  (-1)^{x+\frac{y-z-1}{3}}\cdot 2\cdot 6^{\frac{y+2z-1}{3}}\left(-\frac{4}{3}\right)^{\frac{y-z-1}{3}}, & \text{if }y\equiv z+1\pmod{3},\\
  3\cdot 6^{\frac{y+2z-2}{3}}\left(-\frac{4}{3}\right)^{\frac{y-z+1}{3}}, & \text{if }y\equiv z+2\pmod{3}\text{ and }x+\frac{y-z+1}{3}\text{ is even},\\
  4\cdot 6^{\frac{y+2z-2}{3}}\left(-\frac{4}{3}\right)^{\frac{y-z-2}{3}}, & \text{if }y\equiv z+2\pmod{3}\text{ and }x+\frac{y-z+1}{3}\text{ is odd}.
\end{array}
\right.
$$
The above equations imply that $\langle -1,2,3\rangle\subseteq M\langle 6,-\frac{4}{3}\rangle$, where $M=[-2,4]^{*}$. On the other hand, $6^s\left(-\frac{4}{3}\right)^t=(-1)^t2^{s+2t}3^{s-t}$ for any $s,t\ge 0$. Therefore, $\langle -1,2,3,4\rangle=\langle -1,2,3\rangle=M\langle 6,-\frac{4}{3}\rangle$. Hence, by Lemma~\ref{yl2}, we only need to show that $\langle -1,2,3\rangle=MB$ is a factorization, where
 $$
 B=\left\langle 6,-\frac{4}{3}\right\rangle, \text{ if }\mathrm{ord}_p(6)\text{ is odd,}
 $$
and
 $$
 B=\left\langle 6,-\frac{4}{3}\right\rangle/\left\{1,-1\right\}, \text{ if }\mathrm{ord}_p(6)\text{ is even}.
 $$
Note that if $\mathrm{ord}_p(6)$ is even, then $-1\in\left\langle 6,-\frac{4}{3}\right\rangle$. So by $B=\left\langle 6,-\frac{4}{3}\right\rangle/\{1,-1\}$, we mean that $B$ includes exactly one of $-i,i$  for any $i\in \left\langle 6,-\frac{4}{3}\right\rangle$.
\par
 Since $p\equiv 1\pmod{6}$, we assume that $p=2^a3^bc+1$, where $a,b,c\ge 1$ and $\gcd{(c,6)}=1$. Let $g$ be a primitive root modulo $p$, suppose that
\begin{align*}
2&\equiv g^{2^{u_1}3^{v_1}r_1}\pmod{p},\\
 3&\equiv g^{2^{u_2}3^{v_2}r_2}\pmod{p},\\
 -1&\equiv g^{2^{a-1}3^{b}c}\pmod{p},
\end{align*}
 where $u_1\,u_2,v_1,v_2\ge 0$, $r_1,r_2\ge 1$, $2\nmid r_1r_2,\text{ }3\nmid r_1r_2$, and $2^{u_1}3^{v_1}r_1,2^{u_2}3^{v_2}r_2<p-1$.\par
\par
Note that
$$
\mathrm{ind}_g(4)\equiv 2\times \mathrm{ind}_g(2)\pmod{p-1}
$$
and
$$
\mathrm{ind}_g(-\frac{3}{4})\equiv  2^{u_2}3^{v_2}r_2-\mathrm{ind}_g(4)+2^{a-1}3^{b}c \pmod{p-1},
$$
then
$\mathrm{ord}_p(-\frac{3}{4})$ is odd if and only if
\begin{align}\label{eq2}
 2^{u_2}3^{v_2}r_2-2^{u_1+1}3^{v_1}r_1+2^{a-1}3^{b}c\equiv 0\pmod{2^a}.
\end{align}
By equation~(\ref{eq2}), if $\min\{u_1+1,u_2\}\ge a$, then $2^a\mid 2^{a-1}3^bc$, which is impossible. Similarly, if $\max\{u_1+1,u_2\}\ge a$, then $\min\{u_1+1,u_2\}= a-1$. And if $\max\{u_1+1,u_2\}\le a-1$, then $u_1+1=u_2\le a-2$. Therefore, there are three cases for the possible values of $u_1$ and $u_2$:
\begin{equation}\label{equf}
  \begin{cases}
    u_2=a-1, & \mbox{if } u_1\ge a-1; \\
    u_2\ge a, & \mbox{if } u_1= a-2; \\
    u_2 =u_1+1, & \mbox{otherwise}.
  \end{cases}
\end{equation}
% \begin{itemize}
%   \item $u_1\ge a-1$, $u_2=a-1$,
%   \item $u_1= a-2$, $u_2\ge a$,
%\item $u_1\le a-3$, $u_2 =u_1+1$.
% \end{itemize}

 Since $6=2\times 3$ and $8=2^3$, we have
$$
6\equiv g^{2^{u_1}3^{v_1}r_1+2^{u_2}3^{v_2}r_2}\pmod{p}
$$
and
$$
8\equiv g^{2^{u_1}3^{v_1+1}r_1}\pmod{p}.
$$
Let $d=\gcd\left(2^{u_1}3^{v_1}r_1+2^{u_2}3^{v_2}r_2,2^{u_1}3^{v_1+1}r_1\right)$. So $\langle 6,8\rangle=\langle g^{2^{u_1}3^{v_1}r_1+2^{u_2}3^{v_2}r_2},g^{2^{u_1}3^{v_1+1}r_1}\rangle =\langle g^d\rangle$. We will frequently use this fact in the rest of this proof.
\begin{clm}\label{claim1}
$v_1=v_2$.
\end{clm}
\noindent{\it Proof of Claim~\ref{claim1}}: We split the proof into four cases.

If $u_1\ge u_2$ and $v_1>v_2$, then
$$
d=2^{u_2}3^{v_2}\gcd\left(2^{u_1-u_2}3^{v_1-v_2}r_1+r_2,2^{u_1-u_2}3^{v_1-v_2+1}r_1\right).
$$
Let $r=\gcd\left(2^{u_1-u_2}3^{v_1-v_2}r_1+r_2,2^{u_1-u_2}3^{v_1-v_2+1}r_1\right)$. It is easy to see that $2\nmid r$ and $3\nmid r$, so $r\mid r_1$. It follows that $d\mid 2^{u_2}3^{v_2}r_1$ and thus $d\mid 2^{u_1}3^{v_1}r_1$. Therefore, $2\in \langle 6,8\rangle$, which is a contradiction.

If $u_1\ge u_2$ and $v_1<v_2$, then
$$
d=2^{u_2}3^{v_1}\gcd\left(2^{u_1-u_2}r_1+3^{v_2-v_1}r_2,2^{u_1-u_2}3r_1\right).
$$
If $u_1< u_2$ and $v_1>v_2$, then
$$
d=2^{u_1}3^{v_2}\gcd\left(3^{v_1-v_2}r_1+2^{u_2-u_1}r_2,3^{v_1-v_2+1}r_1\right).
$$
If $u_1< u_2$ and $v_1<v_2$, then
$$
d=2^{u_1}3^{v_1}\gcd\left(r_1+2^{u_2-u_1}3^{v_2-v_1}r_2,3r_1\right).
$$
Similar to the first case, each of these three cases implies that $2\in \langle 6,8\rangle$, which is a contradiction.
This completes the proof of Claim \ref{claim1}.

\begin{clm}\label{claim2}
$v_1=v_2\le b-1$.
\end{clm}
\noindent{\it Proof of Claim~\ref{claim2}}: By computing, we have
\begin{align*}
|\left<-1,2,3\right>|&=\frac{p-1}{\gcd\left(\mathrm{ind}_g(-1),\mathrm{ind}_g(2),\mathrm{ind}_g(3),p-1\right)}\\
&=\frac{p-1}{\gcd\left(2^{u_1}3^{v_1}r_1,2^{u_2}3^{v_2}r_2,2^{a-1}3^bc\right)},
\end{align*}
and
\begin{align*}
|\left<6,8\right>|&=\frac{p-1}{\gcd(d,p-1)}\\
&=\frac{p-1}{\gcd\left(2^{u_1}3^{v_1}r_1+2^{u_2}3^{v_2}r_2,2^{u_1}3^{v_1+1}r_1,2^a3^bc\right)}.
\end{align*}
By Claim~\ref{claim1}, we have $v_1=v_2$. We prove the claim by contradiction. If $v=v_1=v_2\ge b$. Then
\begin{align*}
  \frac{|\left<-1,2,3\right>|}{|\left<6,8\right>|}&=\frac{\gcd\left(2^{u_1}3^vr_1
  +2^{u_2}3^vr_2,2^{u_1}3^{v+1}r_1,2^a3^bc\right)}{\gcd\left(2^{u_1}3^vr_1,2^{u_2}3^vr_2,2^{a-1}3^bc\right)}\\
  &=\frac{\gcd\left(2^{u_1}3^{v-b}r_1
  +2^{u_2}3^{v-b}r_2,2^{u_1}3^{v-b+1}r_1,2^ac\right)}{\gcd\left(2^{u_1}3^{v-b}r_1,2^{u_2}3^{v-b}r_2,2^{a-1}c\right)}.
\end{align*}

If $u_1+1=u_2\le a-2$, then
\begin{align*}
  \frac{|\left<-1,2,3\right>|}{|\left<6,8\right>|}&=\frac{\gcd\left(3^{v-b}r_1
  +2\cdot 3^{v-b}r_2,3^{v-b+1}r_1,2^{a-u_1}c\right)}{\gcd\left(3^{v-b}r_1,2\cdot 3^{v-b}r_2,2^{a-1-u_1}c\right)}\\
  &=\frac{\gcd\left(r_1+r_2,r_1,c\right)}{\gcd\left(r_1,r_2,c\right)}=1.
\end{align*}
On the other hand, it is easy to see that $\langle 6,8\rangle\subseteq\langle -1,2,3\rangle$.
Hence $\langle 6,8\rangle=\langle -1,2,3\rangle$, which contradicts the fact that $2\notin\langle 6,8\rangle$.

Similarly, if $u_1\ge a-1, u_2=a-1$, then
\begin{align*}
  \frac{|\left<-1,2,3\right>|}{|\left<6,8\right>|}&=\frac{\gcd\left(2^{u_1-u_2}3^{v-b}r_1
  +3^{v-b}r_2,2^{u_1-u_2}3^{v-b+1}r_1,2c\right)}{\gcd\left(2^{u_1-u_2}3^{v-b}r_1,3^{v-b}r_2,c\right)}=1.
\end{align*}
If $u_1= a-2, u_2\ge a$, then
\begin{align*}
  \frac{|\left<-1,2,3\right>|}{|\left<6,8\right>|}&=\frac{\gcd\left(3^{v-b}r_1
  +2^{u_2-u_1}3^{v-b}r_2,3^{v-b+1}r_1,4c\right)}{\gcd\left(3^{v-b}r_1,2^{u_2-u_1}3^{v-b}r_2,2c\right)}=1.
\end{align*}
Hence for both cases $\langle 6,8\rangle=\langle -1,2,3\rangle$, which contradicts the fact that $2\notin\langle 6,8\rangle$.
This completes the proof of Claim \ref{claim2}.

Hence from now on, we let $v=v_1=v_2\le b-1$.
Note that
 $$\mathrm{ord}_p(6)=\frac{p-1}{\gcd(2^{u_1}3^{v}r_1+2^{u_2}3^{v}r_2,p-1)},$$
which  is odd if and only if $2^{u_1}3^{v}r_1+2^{u_2}3^{v}r_2\equiv 0\pmod{2^a}$. We can also compute  that
\begin{align}
 \frac{|\left<-1,2,3\right>|}{|\left<6,-\frac{3}{4}\right>|}&=\frac{\gcd\left(2^{u_1}3^{v}r_1
 +2^{u_2}3^{v}r_2,2^{u_2}3^{v}r_2-2^{u_1+1}3^{v}r_1+2^{a-1}3^{b}c,2^{a}3^{b}c\right)}{\gcd\left(2^{u_1}3^{v}r_1,
 2^{u_2}3^{v}r_2,2^{a-1}3^{b}c\right)}\label{eqd}\\
 &=\frac{\gcd\left(2^{u_1}r_1
 +2^{u_2}r_2,2^{u_2}r_2-2^{u_1+1}r_1+2^{a-1}3^{b-v}c,2^{a}3^{b-v}c\right)}{\gcd\left(2^{u_1}r_1,
 2^{u_2}r_2,2^{a-1}3^{b-v}c\right)}.\label{eqe}
\end{align}

Now we divide our proof into two cases.

\noindent\textbf{Case 1: $\bm{\mathrm{ord}_p(6)}$ is odd.} \par
For this case, we have $2^{u_1}3^{v}r_1+2^{u_2}3^{v}r_2\equiv 0\pmod{2^a}$. So we get that $u_1,u_2\ge a$ or $u_1=u_2\le a-1$. But by (\ref{equf}), it forces that $u_1=u_2=a-1$. Thus (\ref{eqe}) becomes
\begin{align*}
 \frac{|\left<-1,2,3\right>|}{|\left<6,-\frac{3}{4}\right>|}&=\frac{\gcd\left(r_1
 +r_2,r_2+3^{b-v}c-2r_1,2 \times 3^{b-v}c\right)}{\gcd\left(r_1,
 r_2,3^{b-v}c\right)}\\
 &=\frac{\gcd\left(r_1
 +r_2,r_2+3^{b-v}c-2r_1,2\times 3^{b-v}c\right)}{\gcd\left(r_1,
 r_2,c\right)}.
\end{align*}
Since $u_1=u_2=a-1$, we have
$$
d=2^{u_2}3^{v}\gcd\left(r_1+r_2,3r_1\right).
$$
\par
If $3\nmid\gcd\left(r_1+r_2,3r_1\right)$, then $d\mid\mathrm{ind}_g(2)$, and hence $2\in\langle 6,8\rangle$, which is a contradiction. Thus $r_1+r_2\equiv 0\pmod{3}$. Now it is easy to see that $2\mid \gcd\left(r_1
 +r_2,r_2+3^{b-v}c-2r_1,2\cdot 3^{b-v}c\right)$ and $3\mid\gcd\left(r_1
 +r_2,r_2+3^{b-v}c-2r_1,2\cdot 3^{b-v}c\right)$. Therefore,
$$
 \frac{\gcd\left(r_1
 +r_2,r_2+3^{b-v}q-2r_1,2\cdot 3^{b-v}c\right)}{\gcd\left(r_1,
 r_2,c\right)}\ge 6,
 $$
since $2\nmid r_1r_2$, $3\nmid r_1r_2$ and $\gcd(c,6)=1$.
This leads to $\frac{\left|\left<-1,2,3\right>\right|}{|\left<6,-\frac{3}{4}\right>|}\ge 6$. On the other hand, $$|\left<-1,2,3\right>|=\left|M\left<6,-\frac{3}{4}\right>\right|\le|M|\left|\left<6,-\frac{3}{4}\right>\right|= 6\left|\left<6,-\frac{3}{4}\right>\right|,$$
therefore, $\left<-1,2,3\right>=MB$ is a factorization.

\noindent\textbf{Case 2: $\bm{\mathrm{ord}_p(6)}$ is even.} \par
For this case, we only need to prove that
$$
\frac{|\left<-1,2,3\right>|}{|\left<6,-\frac{3}{4}\right>|}\ge 3.
$$
We divide our proof into three subcases.

\noindent\textbf{Subcase 1: $u_1\ge a-1$, $u_2=a-1$.}

For this case, we have
$$
d=2^{u_2}3^{v}\gcd\left(2^{u_1-u_2}r_1+r_2,2^{u_1-u_2}3r_1\right).
$$
If $3\nmid\gcd\left(2^{u_1-u_2}r_1+r_2,2^{u_1-u_2}3r_1\right)$, then $2\in\left<6,8\right>$, which is a contradiction. So
$$r_2+2^{u_1-u_2}r_1\equiv 0\pmod{3}\text{ and hence }r_2-2^{u_1-u_2+1}r_1\equiv 0\pmod{3}.$$
Then from  (\ref{eqe}) we have
$$
\frac{|\left<-1,2,3\right>|}{|\left<6,-\frac{3}{4}\right>|}\ge 3.
$$

\noindent\textbf{Subcase 2: $u_1=a-2$, $u_2\ge a$.}

For this case, we have
$$
d=2^{u_1}3^{v}\gcd\left(r_1+2^{u_2-u_1}r_2,3r_1\right).
$$
If $3\nmid\gcd\left(r_1+2^{u_2-u_1}r_2,3r_1\right)$, then $2\in\left<6,8\right>$, which is a contradiction. So
$$r_1+2^{u_2-u_1}r_2\equiv 0\pmod{3}\text{ and hence }2^{u_2-u_1}r_2-2r_1\equiv 0\pmod{3}.$$
Then from  (\ref{eqe})  we have
$$
\frac{|\left<-1,2,3\right>|}{|\left<6,-\frac{3}{4}\right>|}\ge 3.
$$

\noindent\textbf{Subcase 3: $u_1+1=u_2\le a-2$.}

For this case, we have
$$
d=2^{u_1}3^{v}\gcd\left(r_1+2r_2,3r_1\right).
$$
If $3\nmid\gcd\left(r_1+2r_2,3r_1\right)$, then $2\in\left<6,8\right>$, which is a contradiction. So
$$r_1+2r_2\equiv 0\pmod{3}\text{ and hence }r_2-r_1\equiv 0\pmod{3}.$$
Then from  (\ref{eqe})  we have
$$
\frac{|\left<-1,2,3\right>|}{|\left<6,-\frac{3}{4}\right>|}\ge 3.
$$
\end{pf}

\begin{thm}\label{thm1}
   Let $p\equiv 1\pmod{8}$ be a prime, then there exists a nonsingular perfect $B[-4,4](p)$ set if and only if $\pm 4\notin\langle 6,16\rangle$.
\end{thm}
\begin{pf}
First, suppose $B$ is a nonsingular perfect $B[-4,4](p)$ set. Let $\pm B=B\cup (-B)$, $M=\{\pm 1,\pm 2,\pm 3,\pm 4\}$ and $M^{\prime}=\{1,2,3,4\}$. Since $B$ is a perfect $B[-4,4](p)$ set, then by Lemma~\ref{uniqint}, $|B\cap a M|=1$ for any $a\in\mathbb{Z}_p^{*}$. It's easy to verify that $|B\cap aM|=1$ is equivalent to $|(\pm B)\cap aM^{\prime}|=1$. Similarly, $\mathbb{Z}_p^{*}=MB$ is a factorization if and only if $\mathbb{Z}_p^{*}=M^{\prime}(\pm B)$ is a factorization.
\par
Note that if $\mathbb{Z}_p^{*}=MB$ is a factorization, then $\mathbb{Z}_p^{*}=MB^{\prime}$ is also a factorization, where $B^{\prime}=b^{-1}B$ for some $b\in B$. Hence, without loss of generality, we may assume that $1\in\pm B$. If $r\in\pm B$, then from $|\pm B\cap rM^{\prime}|=1$, we have $2r,3r,4r\notin\pm B$; from $|\pm B\cap \frac{1}{2}r M^{\prime}|=1$, we have  $\frac{3}{2}r\notin\pm B$; and from $|\pm B\cap \frac{1}{3}r M^{\prime}|=1$, we have $\frac{2}{3}r,\frac{4}{3}r\notin\pm B$. Note that $6r\in \mathbb{Z}_p^{*}=M^{\prime}(\pm B)$, which can be written as $6r=1\cdot (6r)=2\cdot (3r)=3\cdot (2r)=4\cdot(\frac{3}{2}r)$, but $2r,3r, \frac{3}{2}r\notin\pm B$, so we have $6r\in\pm B$. Since $|\pm B\cap 2rM^{\prime}|=1$, $|\pm B\cap 3rM^{\prime}|=1$ and $|\pm B\cap 4rM^{\prime}|=1$, then $8r,9r,12r\notin\pm B$ and $16r\in\pm B$.
\par
With the observation above and the fact $1\in\pm B$, it is easy to see that $\langle 6,16\rangle\subseteq \pm B$ and
$\left<6,16\right>\cap\{\pm 2,\pm3,\pm 4,\pm 8,\pm\frac{2}{3},\pm\frac{4}{3}\}=\emptyset$, which leads to $\pm 4\notin\left<6,16\right>$.
\par
For the other direction, suppose $\pm 4\notin\langle 6,16\rangle$. Then $ 2, 4, 8\notin\langle 6,16\rangle$ (if $8\in\langle 6,16\rangle$, then $2\in\langle 6,16\rangle$) and $16\in \langle 6,16\rangle$. So the order of $2\langle 6,16\rangle$ in the quotient group $\langle -1,2,3\rangle/\langle 6,16\rangle$ is $4$. Since $3\times 16=8\times 6$, we have $3\langle 6,16\rangle=8\langle 6,16\rangle$. Therefore, the subgroup of $\langle -1,2,3\rangle/\langle 6,16\rangle$ generated by $2\langle 6,16\rangle$ is
$$
\left\langle 2\langle 6,16\rangle \right\rangle=\{\langle 6,16\rangle,2\langle 6,16\rangle,3\langle 6,16\rangle,4\langle 6,16\rangle\}.
$$
In particular, $|\langle -1,2,3\rangle|\ge 4|\langle 6,16\rangle|$.
\begin{claim}
$-1\in\langle 6,16\rangle$.
\end{claim}
\noindent{\it Proof of Claim}:
Since $p\equiv 1\pmod{8}$, we can assume $p=2^bc+1$, where $b, c$ are integers, $b\ge 3$ and $\gcd(c,2)=1$. Let $g$ be a primitive root modulo $p$ and suppose that
\begin{align*}
  2\equiv g^{2^{u_1}r_1}\pmod{p} \text{ and }
  3\equiv g^{2^{u_2}r_2}\pmod{p},
\end{align*}
where $u_1,u_2\ge 0,r_1,r_2\ge 1$ are integers and $2\nmid r_1r_2$. Let $d=\gcd(2^{u_1}r_1+2^{u_2}r_2,2^{u_1+2}r_1)$, then $\langle 6,16\rangle=\langle g^d\rangle$.
\par
If $u_1>u_2$, then $d=2^{u_2}\gcd(2^{u_1-u_2}r_1+r_2,2^{u_1-u_2+2}r_1)=2^{u_2}\gcd(r_1,r_2)$. Now it is easy to see that $d\mid 2^{u_1}r_1$, and $2\in\langle 6,16\rangle$, which is a contradiction. Similarly, if $u_1<u_2$, we can also get $2\in\langle 6,16\rangle$. Therefore, we always have $u_1=u_2$. Now we assume $u=u_1=u_2$. Then $d=2^u\gcd(r_1+r_2,4r_1)$. If $4\nmid (r_1+r_2)$, then $d=2^{u+1}\gcd(r_1,r_2)$. So $d\mid 2^{u+1}r_1$ and $4\in\langle 6,16\rangle$, which is a contradiction. Thus, $4\mid (r_1+r_2)$. Then $d=2^{u+2}\gcd(r_1,r_2)$. If $u\ge b-1$, then
\begin{align*}
  \frac{|\langle -1,2,3\rangle|}{|\langle 6,16\rangle|} & =\frac{\gcd(2^{u+2}r_1,2^{u+2}r_2,2^bc)}{\gcd(2^{u}r_1,2^ur_2,2^{b-1}c)}\\
  &=\frac{\gcd(2^{u-b+3}r_1,2^{u-b+3}r_2,2c)}{\gcd(2^{u-b+1}r_1,2^{u-b+1}r_2,c)}\\
  &=\frac{2\times\gcd(2^{u-b+2}r_1,2^{u-b+2}r_2,c)}{\gcd(r_1,r_2,c)}\\
  &=\frac{2\times\gcd(r_1,r_2,c)}{\gcd(r_1,r_2,c)}=2,
\end{align*}
which contradicts the fact that $|\langle -1,2,3\rangle|\ge 4|\langle 6,16\rangle|$. If $u=b-2$, then $\frac{|\langle -1,2,3\rangle|}{|\langle 6,16\rangle|}=4$. So $\langle -1,2,3\rangle=\{1,2,3,4\}\langle 6,16\rangle$ is a factorization. This means $-\langle 6,16\rangle=i\langle 6,16\rangle$ for some $i\in\{1,2,3,4\}$, that is to say, $-i\in\langle 6,16\rangle$. Note that $\pm 4\notin\langle 6,16\rangle$, so $i\ne 2,3,4$. On the other hand, since $u=b-2$, we have
$$
|\langle 6,16\rangle|=\frac{p-1}{\gcd(d,p-1)}=\frac{c}{\gcd(r_1,r_2,c)}
$$
is an odd number, which means $-1\notin\langle 6,16\rangle$.
Therefore, $u\le b-3$. Now
\begin{align*}
  |\langle 6,16\rangle| & =\frac{p-1}{\gcd(2^{u+2}r_1,2^{u+2}r_2,2^bc)}\\
  &=\frac{2^{b-u-2}c}{\gcd(r_1,r_2,2^{b-u-2}c)}
\end{align*}
is an even number. So $-1\in\langle 6,16\rangle$. This completes the proof of the claim.

\par
Since $-1\in\langle 6,16\rangle$, then $\langle -1,2,3\rangle=\langle 2,6,16\rangle=\{1, 2, 3, 4\}\langle 6,16\rangle$ is a factorization.  Let $a=\gcd\left(\frac{p-1}{2},\mathrm{ind}_g(6),\mathrm{ind}_g(16)\right)$, then $\langle 6,16\rangle=\left\langle g^{a}\right\rangle$.
Let $u\ge 1$ be the smallest integer such that $2^ua\nmid\frac{p-1}{2}$, then $-1\notin \left\langle g^{2^ua}\right\rangle$. Let $S$ be a complete set of coset representatives of $\left\langle g^{2^ua}\right\rangle$ in $\langle 6,16\rangle$. Since $-1\in\langle 6,16\rangle$ and $-1\notin\left\langle g^{2^ua}\right\rangle$, we can choose $S$ such that if $s\in S$, then $-s\in S$. Let $S^{\prime}=\left\{s:\text{ }s\in S\text{ and }0\le\mathrm{ind}_g(s)<\frac{p-1}{2}\right\}$. Then
$$
\langle -1,2,3\rangle=\{\pm 1,\pm 2,\pm 3,\pm 4\}\left(\mathop{\bigcup}_{s\in S^{\prime}}s\left\langle g^{2^ua}\right\rangle\right)
$$
is a factorization. By Lemma \ref{yl2}, there exists a perfect $B[-4,4](p)$ set.
\end{pf}
\begin{rmk}
We note that perfect $B[-4,4](p)$ sets have been considered in \cite{M95} and \cite{wz14} before.
\par
In \cite[Lemma 4.3]{M95}, the author gave an equivalent condition for the existence of a perfect $B[-4,4](p)$ set. But the construction method in the proof of Theorem \ref{thm1} is more explicit and simpler than that in the proof of \cite[Lemma 4.3]{M95}.\par
In \cite[Theorem 1]{wz14}, Tamm gave an equivalent condition for the existence of a perfect $B[-4,4](p)$ set and claimed that a perfect $B[-4,4](p)$ set must be of the form
$$
x_0\cdot\mathcal{F}\cup\cdots \cup x_{\rho-1}\cdot\mathcal{F}.
$$
However, the calculations of $x_0,\ldots,x_{\rho-1}$ and $\mathcal{F}$ make his construction more complicated than ours.
\end{rmk}

 Similar to Theorem~\ref{thm1}, we show the following result, for which we just sketch the proof.
\begin{thm}\label{thm2}
   Let $p\equiv 1\pmod{4}$ be a prime, then there exists a nonsingular perfect $B[0,4](p)$ set if and only if $4\notin\langle 6,16\rangle$.
\end{thm}
\begin{pf}
  First, suppose $B$ is a nonsingular perfect $B[0,4](p)$ set. Let $M=\{1, 2, 3, 4\}$. In the proof of Theorem~\ref{thm1},   taking $M'$ as $M$,  $B'$ as $B$, respectively, and following the same procedure, we get $\langle 6,16\rangle\subseteq B$ and $4\notin\langle 6,16\rangle$.
  \par
  For the other direction, as in the proof of Theorem~\ref{thm1}, the fact $4\notin\langle 6,16\rangle$ implies that
$$
\left\langle 2\langle 6,16\rangle \right\rangle=\{\langle 6,16\rangle,2\langle 6,16\rangle,3\langle 6,16\rangle,4\langle 6,16\rangle\}.
$$
Therefore, $\langle 1,2,3,4\rangle=\langle 2,6,16\rangle=\{1, 2, 3, 4\}\langle 6,16\rangle$ is a factorization. By Lemma \ref{yl2}, there exists a perfect $B[0,4](p)$ set.
\end{pf}

If there exists a nonsingular perfect $B[-2,4](p)$ set ($B[-4,4](p)$ set, or  $B[0,4](p)$ set), then we can construct it explicitly from the proofs of  Lemma~\ref{yl2} and Theorem~\ref{thm0} (Theorem~\ref{thm1}, or Theorem~\ref{thm2}, respectively).
\begin{example}
We give three examples to illustrate how to construct perfect splitter sets by using the above theorems.
\begin{enumerate}
  \item We use the same notations as in the proof of Theorem~\ref{thm1}. Let $p=97$, then $g=5$, $\mathrm{ind}_g(6)=8$, $\mathrm{ind}_g(4)=68$, $\mathrm{ind}_g(-4)=20$,  $\mathrm{ind}_g(16)=40$ and $a=8$. Since $8x\equiv 68\pmod{96}$ has no solution, then $4\notin\langle 6,16\rangle$. Similarly, $-4\notin\langle 6,16\rangle$. So by Theorem~\ref{thm1}, there exists a perfect $B[-4,4](p)$ set. Further,
  $$
  \langle 6,16\rangle=\{1,6,16,22,35,36,61,62,75,81,91,96\}
  $$
and (here $u=2$)
$$
\left\langle g^{2^ua}\right\rangle=\left\langle 5^{32}\right\rangle=\{1,35,61\}.
$$
We can choose $S=\{1,6,91,96\}$ and $S^{\prime}=\{1,6\}$. Furthermore, $T=\{1,5\}$ is a complete set of coset representatives of $\langle -1,2,3\rangle$ in $\mathbb{Z}_{97}^{*}$. Thus, the set
$$
\mathop{\bigcup}_{t\in T}\mathop{\bigcup}_{s\in S^{\prime}}\left\{sti\pmod{97}:\text{ }i\in\left\langle g^{2^ua}\right\rangle\right\}=\{1,5,6,14,16,30,35,61,75,78,80,84\}
$$
is a perfect $B[-4,4](97)$ set. By Theorem~\ref{thm1} and computation, when $p\le 5000$ is a prime, there is a perfect $B[-4,4](p)$ set if and only if $p=97,1873,2161$ and $3457$.
  \item Let $p=139$, then $g=2$, $\mathrm{ind}_g(2)=1$, $\mathrm{ind}_g(6)=42$ and $\mathrm{ind}_g(8)=3$, so $2\notin\langle 6,8\rangle$. It is easy to see that $-\frac{4}{3}=45$ in $\mathbb{Z}_{139}$ and $\mathrm{ord}_p(45)=23$. Therefore, there exists a perfect $B[-2,4](139)$ set. In this case, $\mathrm{ord}_p(6)=23$ is odd, $\mathrm{ind}_g(45)=30$ and $\gcd(\mathrm{ind}_g(6),\mathrm{ind}_g(45))=6$. By the proof of Theorem~\ref{thm0}, the set
$$
\langle 6,45\rangle=\{2^{6i}\pmod{139}:\text{ }0\le i\le 22\}
$$
is a perfect $B[-2,4](139)$ set.
  \item Let $p=181$, then $g=2$, $\mathrm{ind}_g(2)=1$, $\mathrm{ind}_g(6)=57$ and $\mathrm{ind}_g(8)=3$, so $2\notin\langle 6,8\rangle$. It is easy to see that $-\frac{4}{3}=59$ and $\mathrm{ord}_p(59)=5$. Therefore, there exists a perfect $B[-2,4](181)$ set. In this case, $\mathrm{ord}_p(6)=60$ is even, $\mathrm{ind}_g(59)=36$ and $\gcd(\mathrm{ind}_g(6),\mathrm{ind}_g(59))=3$. Then $\langle 6,59\rangle=\{2^{3i}\pmod{181}:\text{ }0\le i\le 59\}$. By the proof of Theorem~\ref{thm0}, the set
$$
\{2^{3i}\pmod{181}:\text{ }0\le i\le 29\}
$$
is a perfect $B[-2,4](181)$ set. By Theorem~\ref{thm0}, for primes $p\le 1000$, apart from the $10$ constructions in \cite{wz2}, perfect $B[-2,4](p)$ sets exist only when $p=181,313,421,541,919$ and $937$.
\end{enumerate}

\end{example}

\subsection{Simpler characterizations for special cases}
\indent When $\gcd\left(\frac{p-1}{k_1+k_2},k_1+k_2\right)=1$, we are able to give a much simpler characterization for the existence of perfect splitter sets. Before stating our results, we need some useful lemmas.
\begin{lem}\label{yl3}\textup{\cite[Theorem 7.1]{bk1}}
Let $m$ and $n$ be relatively prime positive integers. If $A=\{a_1,\ldots,a_m\}$ and $B=\{b_1,\ldots,b_n\}$ are sets of integers such that their sum set
$$
A+B:=\{a_i+b_j:\text{ }1\le i\le m,1\le j\le n\}
$$
is a complete set of representatives modulo $mn$, then $A$ is a complete set of residues modulo $m$ and $B$ is a complete set of residues modulo $n$.
\end{lem}

\begin{lem}\label{yl4}
Let $ k_2\geq k_1\geq 0$ be integers, and let $p$ be a prime such that $p\equiv 1 \pmod{k_1+k_2}$ and $\gcd\left(k_1+k_2,\frac{p-1}{k_1+k_2}\right)=1$. Suppose $g$ is a primitive root modulo $p$, and denote $N=\{\mathrm{ind}_g(j):\text{ }j\in[-k_1,k_2]^{*}\}$. Then there exists a nonsingular perfect $B[-k_1,k_2](p)$ set if and only if $N$ is a complete set of residues modulo $k_1+k_2$.
\end{lem}
\begin{pf}
Let $B$ be a nonsingular perfect $B[-k_1,k_2](p)$ set, and $A=\{\mathrm{ind}_g(b):\text{ }b\in B\}$. Then $\mathbb{Z}_{p-1}=N+A$ is a factorization. Since $\gcd\left(k_1+k_2,\frac{p-1}{k_1+k_2}\right)=1$, it follows from Lemma \ref{yl3} that $N$ is a complete set of residues modulo $k_1+k_2$.\par
The other direction follows from \cite[Theorem 3]{wz2}.
\end{pf}

We will apply Lemma~\ref{yl4} to the cases when $(k_1,k_2)\in \{(2,4),(0,4),(4,4)\}$. Note that $p\equiv 1\pmod{6}$ and $\gcd\left(6,\frac{p-1}{6}\right)=1$ if and only if $p\equiv 7,31\pmod{36}$;  $\gcd\left(4,\frac{p-1}{4}\right)=1$ is equivalent to that $p\equiv 5\pmod{8}$; and $\gcd\left(8,\frac{p-1}{8}\right)=1$ is equivalent to that $p\equiv 9\pmod{16}$.
\begin{thm}\label{dl4}
\begin{enumerate}
\item Let $p\equiv 7, 31\pmod{36}$ be a prime. Then there exists a nonsingular perfect $B[-2,4](p)$ set if and only if $6$ is a cubic residue in $\mathbb{Z}_p$ and $2,3$ are not cubic residues.
\item Let $p\equiv 5\pmod{8}$ be a prime. Then there exists a nonsingular perfect $B[0,4](p)$ set if and only if $6$ is a quartic residue modulo $p$.
\item Let $p\equiv 9\pmod{16}$ be a prime. Then there does not exist a nonsingular perfect $B[-4,4](p)$ set.
\end{enumerate}
\end{thm}
\begin{pf}
Assume that $g$ is a primitive root modulo $p$.
\par
1).
Suppose there exists a nonsingular perfect $B[-2,4](p)$ set. By Lemma \ref{yl4}, $N=\{\mathrm{ind}_{g}(j)\text{ }\pmod{6}:\text{ }j\in[-2,4]^{*}\}=\mathbb{Z}_{6}$. Since $\mathrm{ind}_g(-j)\equiv \mathrm{ind}_g(j)+\frac{p-1}{2}\pmod{p-1}$, we have $\mathrm{ind}_g(-j)\equiv \mathrm{ind}_g(j)+3\pmod{6}$. The possible values of $\mathrm{ind}_g(j)$ modulo $6$ are listed below.\par
\begin{center}
\begin{tabular}{|c|c|c|c|c|}
  \hline
   & case 1 & case 2 & case 3 & case 4 \\
   \hline
  $\mathrm{ind}_{g}(1)\text{ }\pmod{6}$ & $0$ & $0$ & $0$ & $0$ \\
  \hline
  $\mathrm{ind}_{g}(-1)\text{ }\pmod{6}$ & $3$ & $3$ & $3$ & $3$ \\
  \hline
  $\mathrm{ind}_{g}(2)\text{ }\pmod{6}$ & $1$ & $2$ & $4$ & $5$ \\
  \hline
  $\mathrm{ind}_{g}(-2)\text{ }\pmod{6}$ & $4$ & $5$ & $1$ & $2$ \\
  \hline
  $\mathrm{ind}_{g}(3)\text{ }\pmod{6}$ & $5$ & $1$ & $5$ & $1$ \\
  \hline
  $\mathrm{ind}_{g}(4)\text{ }\pmod{6}$ & $2$ & $4$ & $2$ & $4$ \\
  \hline
\end{tabular}\par
\end{center}
Therefore integers $2$ and $3$ can not be cubic residues modulo $p$, and $6$ must be a cubic residue whichever the case is.
\par
For the other direction, suppose $\mathrm{ind}_g(2)=x$ and $\mathrm{ind}_g(3)=y$. Then $\mathrm{ind}_g(6)\equiv x+y\pmod{p-1}$. The fact that $6$ is a cubic residue implies that $x+y\equiv 0\pmod{3}$. Further, the fact that $2$ and $3$ are not cubic residues implies that $\{x,y\}\equiv\{1,2\}\text{ or }\{1,5\}\text{ or }\{2,4\}\text{ or }\{4,5\}\pmod{6}$. Since $p\equiv 7,31\pmod{36}$, $3$ is not a quadratic residue in $\mathbb{Z}_p$ \cite[page 55]{bk2}. Combining all these observations, we have the only four cases for the values of $x$ and $y$. For any case, it is easy to see that $\{\mathrm{ind}_{g}(j)\text{ }\pmod{6}:\text{ }j\in[-2,4]^{*}\}=\mathbb{Z}_{6}$. The proof is complete by Lemma \ref{yl4}.\par
\begin{center}
\begin{tabular}{|c|c|c|c|c|}
  \hline
   & case 1 & case 2 & case 3 & case 4 \\
   \hline
  $x\pmod{6} $ & $2$ & $1$ & $5$ & $4$ \\
  \hline
  $y\pmod{6} $ & $1$ & $5$ & $1$ & $5$ \\
  \hline
\end{tabular}\par
\end{center}

2). Suppose there exists a nonsingular perfect $B[0,4](p)$ set. Then $\{\mathrm{ind}_{g}(j)\text{ }\pmod{4}:\text{ }j\in[0,4]^{*}\}=\mathbb{Z}_{4}$ by Lemma \ref{yl4}. Since $p\equiv 5\pmod{8}$, then $2$ is not a quadratic residue modulo $p$ \cite[Proposition 5.1.3]{bk2}. Note also that $\mathrm{ind}_g(4)\equiv 2\times\mathrm{ind}_g(2)\pmod{4}$. Therefore, there are only two case: $\mathrm{ind}_g(2)\equiv 1\pmod{4}$, $\mathrm{ind}_g(3)\equiv 3\pmod{4}$, $\mathrm{ind}_g(4)\equiv 2\pmod{4}$ or $\mathrm{ind}_g(2)\equiv 3\pmod{4}$, $\mathrm{ind}_g(3)\equiv 1\pmod{4}$, $\mathrm{ind}_g(4)\equiv 2\pmod{4}$. In both cases, we have $\mathrm{ind}_g(6)\equiv \mathrm{ind}_g(2)+\mathrm{ind}_g(3)\equiv 0\pmod{4}$, that is, $6$ is a quartic residue modulo $p$.
\par
For the other direction, suppose $\mathrm{ind}_g(2)=x$ and $\mathrm{ind}_g(3)=y$.  Then the fact that $6$ is a quartic residue modulo $p$ implies that  $\mathrm{ind}_g(6)\equiv x+y\equiv 0\pmod{4}$. We also have $x\equiv 1\text{ or }3\pmod{4}$ since $2$ is not a quadratic residue modulo $p$. Thus, we have two cases.
\begin{center}
\begin{tabular}{|c|c|c|}
  \hline
   & case 1 & case 2\\
   \hline
  $\mathrm{ind}_{g}(1)\text{ }\pmod{4}$ & $0$ & $0$ \\
  \hline
  $\mathrm{ind}_{g}(2)\text{ }\pmod{4}$ & $1$ & $3$\\
  \hline
  $\mathrm{ind}_{g}(3)\text{ }\pmod{4}$ & $3$ & $1$\\
  \hline
  $\mathrm{ind}_{g}(4)\text{ }\pmod{4}$ & $2$ & $2$\\
  \hline
\end{tabular}
\end{center}
For any case, it is easy to see that $\{\mathrm{ind}_{g}(j)\text{ }\pmod{4}:\text{ }j\in[0,4]^{*}\}=\mathbb{Z}_{4}$. The proof is complete.

3). Since  $\frac{p-1}{2}\equiv 4\pmod{8}$, we always have $\mathrm{ind}_g(1)\equiv 0\pmod{8}$ and $\mathrm{ind}_g(-1)\equiv 4\pmod{8}$. Since $p\equiv 1\pmod{8}$, $2$ is a quadratic residue modulo $p$. There are four cases for $\mathrm{ind}_g(2)$:
\begin{itemize}
  \item $\mathrm{ind}_g(2)\equiv 0\pmod{8}$;
  \item $\mathrm{ind}_g(2)\equiv 2\pmod{8}$, then $\mathrm{ind}_g(4)\equiv 4\pmod{8}$;
  \item $\mathrm{ind}_g(2)\equiv 4\pmod{8}$, then $\mathrm{ind}_g(-2)\equiv 0\pmod{8}$; and
  \item $\mathrm{ind}_g(2)\equiv 6\pmod{8}$, then $\mathrm{ind}_g(4)\equiv 4\pmod{8}$.
\end{itemize}
For any case, it is impossible to have $\{\mathrm{ind}_{g}(j)\text{ }\pmod{8}:\text{ }j\in[-4,4]^{*}\}=\mathbb{Z}_{8}$. Hence there does not exist a perfect $B[-4,4](p)$ set.
\end{pf}
\par
~\\
\indent For the existence of nonsingular perfect $B[-2,4](p)$ sets, we can give another characterization from number theory. In the following discussion, all the undefined terminologies can be found in \cite{bk2}.
\par
Let $\omega=\frac{-1+\sqrt{-3}}{2}$. Suppose $p\equiv 1\pmod{6}$, then we can assume $p=\pi\bar{\pi}$, where $\pi=3m-1+3n\omega$ is a primary prime in the ring $\mathbb{Z}[\omega]$, and $\bar{\pi}$ is the complex conjugate of $\pi$. By the cubic reciprocity and \cite[Chapter 9, Exercise 5]{bk2}, we have
$$
\chi_{\pi}(2)=\chi_{2}(\pi)\equiv\pi\pmod{2}\text{\ \ \ and\ \ }\chi_{\pi}(3)=\omega^{2n}.
$$
Notice that $\chi_{\pi}(2),\chi_{\pi}(3)\ne 1$, as that $2$ and $3$ are not cubic residues modulo $p$ (and therefore modulo $\pi$). Thus, $6$ is a cubic residue modulo $p$ if and only if
$$
\left\{
\begin{array}{l}
\chi_{\pi}(2)=\omega\\
\chi_{\pi}(3)=\omega^2
\end{array}
\right.
\text{ or }
\left\{
\begin{array}{l}
\chi_{\pi}(2)=\omega^2\\
\chi_{\pi}(3)=\omega,
\end{array}
\right.
$$
and hence if and only if
\begin{equation}\label{eq1}
\left\{
\begin{array}{l}
m\text{ is odd, } n\text{ is odd}\\
n\equiv 1\pmod{3}
\end{array}
\right.
\text{ or }
\left\{
\begin{array}{l}
m\text{ is even, } n\text{ is odd}\\
n\equiv 2\pmod{3}.
\end{array}
\right.
\end{equation}
\indent For the first condition in~(\ref{eq1}), let $m=2k+1$ for some integer $k$. Since $n$ is odd and $n\equiv 1\pmod{3}$, then $n$ can only be of the form $6l+1$ for some integer $l$. In this case, $p=\pi\bar{\pi}=36k^2-108kl+324l^2+6k+72l+7$, then $\frac{p-1}{6}\equiv k+1\pmod{6}$. So $\gcd(\frac{p-1}{6},6)=1$ if and only if $k\equiv 0\text{ or }4\pmod{6}$, that is $m\equiv 1\text{ or }9\pmod{12}$.
\par
For the second condition in~(\ref{eq1}), let $m=2k$ for some integer $k$. Since $n$ is odd and $n\equiv 2\pmod{3}$, then $n$ can only be of the form $6l+5$ for some integer $l$. In this case, $p=\pi\bar{\pi}=36k^2-108kl+324l^2-102k+558l+241$, then $\frac{p-1}{6}\equiv k+3l+4\pmod{6}$.  So $\gcd(\frac{p-1}{6},6)=1$ if and only if $k+3l\equiv 1\text{ or }3\pmod{6}$.
\par
Thus, we have the following corollary.
\begin{cor}\label{corollary1}
  Let $p\equiv 1\pmod{6}$ be a prime and $\gcd(\frac{p-1}{6},6)=1$. Then there exists a nonsingular perfect $B[-2,4](p)$ set if and only if there exist $k,l\in\mathbb{Z}$, such that one of the following three conditions holds:
\begin{enumerate}
  \item $p=1296k^2-648kl+324l^2+36k+72l+7$. This case corresponds to the first condition in~(\ref{eq1}) and $m\equiv 1\pmod{12}$.
  \item $p=1296k^2-648kl+324l^2+1764k-360l+607$. This case corresponds to the first condition in~(\ref{eq1}) and $m\equiv 9\pmod{12}$.
  \item $p=36k^2-108kl+324l^2-102k+558l+241$ and $k+3l\equiv 1\text{ or }3\pmod{6}$. This case corresponds to the second condition in~(\ref{eq1}).
\end{enumerate}
\end{cor}
\par
\begin{example}
We give some examples from Corollary~\ref{corollary1}.
  \begin{enumerate}
    \item Let $k,l$ range from $-100$ to $100$. The eight smallest primes of the form $p=1296k^2-648kl+324l^2+36k+72l+7$ are listed in Table \ref{s1}.
    \begin{table}[htbp]
      \centering
      \caption{}\label{s1}%$p=1296k^2-648kl+324l^2+36k+72l+7$
      \begin{tabular}{|c|c|c|c|c|c|c|c|c|}
        \hline
        $p$ &$7$&$1087$&$1123$&$1447$&$1483$&$2239$&$2311$&$2707$\\
        \hline
        $k$ &$0$&$1$&$-1$&$0$&$1$&$1$&$-1$&$0$\\
        \hline
        $l$ &$0$&$1$&$-2$&$2$&$2$&$-1$&$1$&$-3$\\
        \hline
      \end{tabular}
    \end{table}
    In particular, if we let $l=0$, then $p=1296k^2+36k+7$. Bunyakovsky's conjecture \cite{web1}, which has not been proved yet, suggests that there are infinitely many such primes.
    \item Let $k,l$ range from $-100$ to $100$. The eight smallest primes of the form $p=1296k^2-648kl+324l^2+1764k-360l+607$ are listed in Table \ref{s2}.
    \begin{table}[htbp]
      \centering
      \caption{}\label{s2}%$p=1296k^2-648kl+324l^2+1764k-360l+607$
      \begin{tabular}{|c|c|c|c|c|c|c|c|c|}
        \hline
        $p$ &$139$&$571$&$607$&$751$&$859$&$1291$&$2011$&$2371$\\
        \hline
        $k$ &$-1$&$0$&$0$&$-1$&$-1$&$0$&$-1$&$-2$\\
        \hline
        $l$ &$0$&$1$&$0$&$1$&$-2$&$-1$&$2$&$-3$\\
        \hline
      \end{tabular}
    \end{table}
  \end{enumerate}
\end{example}

\section{Constructions of quasi-perfect splitter sets}\label{sec4}
In this section, we provide four new constructions of quasi-perfect splitter sets.
\subsection{Quasi-perfect $B[0,k](m)$ sets}
\begin{thm}\label{dl5}
  Let $k,m$ be positive integers such that $\gcd(m,k!)=1$. Let $a=(-k)^{-1}\pmod{m}$. Then
  $$
  B=\{ik+1:\text{ }i\in[0,m-1]\text{ and }i\neq a\}
  $$
  is a quasi-perfect $B[0,k](km)$ set.
\end{thm}
\begin{pf}
  Suppose $r(ik+1)\equiv 0\pmod{km}$, where $r\in[1,k]$ and $i\in[0,m-1]\setminus\{a\}$. Since $ik+1\not\equiv 0 \pmod k$, then $r\equiv 0\pmod{k}$, and hence $r=k$. This implies that $ik+1\equiv 0\pmod{m}$, which contradicts the fact that $i\not\equiv(-k)^{-1}\pmod{m}$.

Suppose $r(ik+1)\equiv s(jk+1)\pmod{km}$, where $r,s\in[1,k]$ and $i,j\in[0,m-1]\setminus\{a\}$. Then $r\equiv s\pmod{k}$, and so $r=s$. This implies that $rik\equiv rjk\pmod{km}$, and so $ri\equiv rj\pmod{m}$. Note that $\gcd(m,k!)=1$, then $i\equiv j\pmod{m}$, and so $i=j$.
\par
Combing the above analysis, we see that $B$ is a $B[0,k](km)$ set of size $ m-1=\left\lfloor\frac{km-1}{k}\right\rfloor$.
\end{pf}
\begin{example}
\begin{enumerate}
  \item Let $k=5$ and $m=7$. By Theorem \ref{dl5}, the set
  $$\{1,6,11,16,26,31\}$$
  is a quasi-perfect $B[0,5](35)$ set.
  \item Let $k=6$ and $m=7$. By Theorem \ref{dl5}, the set
  $$\{1,13,19,25,31,37\}$$
  is a quasi-perfect $B[0,6](42)$ set.
\end{enumerate}
\end{example}
\begin{rmk}
  It is easy to see that Theorem \ref{dl5} is a generalization of \cite[Theorem 1]{wz7}, and the above examples cannot be obtained by Theorem 1 of \cite{wz7}. Moreover, Theorem \ref{dl5} shows that, for any integer $k$, there exists a quasi-perfect $B[0,k](km)$
set for all positive integers $m$ whose prime factors are all greater than k.
\end{rmk}

\subsection{Quasi-perfect $B[-k,k](m)$ sets}
\begin{thm}\label{dl6}
 Let $k>0$ be an integer, and $p$ be a prime such that $k<p<2k$. Then
  $$
  B=\{k+1\}\cup\{1+(2k+2)i:\text{ }i\in[0,p-1]\}
  $$
  is a quasi-perfect $B[-k,k](p(2k+2))$ set.
\end{thm}
\begin{pf}
Suppose $r(k+1)\equiv s(k+1)\pmod{p(2k+2)}$, where $r,s\in[-k,k]^{*}$. Then $r\equiv s\pmod{2p}$, and so $r=s$.
\par
Suppose $r(k+1)\equiv s(1+(2k+2)i)\pmod{p(2k+2)}$, where $r,s\in[-k,k]^{*}$ and $i\in[0,p-1]$. Then $s\equiv 0\pmod{k+1}$, which is a contradiction.
\par
Suppose $r(1+(2k+2)i)\equiv s(1+(2k+2)j)\pmod{p(2k+2)}$, where $r,s\in[-k,k]^{*}$ and $i,j\in[0,p-1]$. Then $r\equiv s\pmod{2k+2}$, and so $r=s$. This implies $r(2k+2)i\equiv r(2k+2)j\pmod{p(2k+2)}$, and so $ri\equiv rj\pmod{p}$. Note that $p>k$ is a prime, we have $\gcd(r,p)=1$. Then $i\equiv j\pmod{p}$, and so $i=j$.
\par
Combing  the above analysis, we see that $B$ is a $B[-k,k](p(2k+2))$ set of size $p+1=\left\lfloor\frac{p(2k+2)-1}{2k}\right\rfloor$.
\end{pf}
\begin{example}
  \begin{enumerate}
  \item\label{ex1} Let $k=3$ and $p=5$. By Theorem \ref{dl6}, the set
  $$\{1,4,9,17,25,33\}$$
  is a quasi-perfect $B[-3,3](40)$ set.
  \item Let $k=4$ and $p=7$. By Theorem \ref{dl6}, the set
  $$\{1,5,11,21,31,41,51,61\}$$
  is a quasi-perfect $B[-4,4](70)$ set.
\end{enumerate}
\end{example}

\begin{thm}\label{dl7}
 Let $k$ be an even integer and $m\ge 1$. For $i=0,1$, let $T_i=\{x:\text{ }x\equiv i\pmod{2},x\in[1,k]\}$, then $|T_i|=\frac{k}{2}$. Suppose that $p\equiv 1\pmod{2^mk}$ is a prime. Let $g$ be a primitive root modulo $p$ such that $g\equiv 1\pmod{2}$. Denote $v:=2^{m-1}k$. If there exists a $2^m$-subset $A\subset \mathbb{Z}_{v}$ such that $\mathbb{Z}_{v}=A+\{\mathrm{ind}_g(x)\pmod{v}:\text{ }x\in T_i\}$ is a factorization for each $i=0,1$, then there exists a quasi-perfect $B[-k,k](2p)$ set.
\end{thm}
\begin{pf} Let $p=2^mkn+1=2vn+1$ be a prime for some $n\ge 1$.
  We claim that the set $\{g^{i+jv}:\text{ }i\in A,j\in[0,n-1]\}$ is a quasi-perfect $B[-k,k](2p)$ set of size $2^mn$.
\par
Suppose that
\begin{equation}\label{eqg}
  sg^{i_{1}+j_{1} v} \equiv l g^{i_{2}+j_{2} v} \text{ }\pmod{2p},
\end{equation}
where $s, l \in[-k, k]^{*}, i_{1}, i_{2} \in A$ and $j_{1}, j_{2} \in[0, n-1]$.
Then
$$
s g^{i_{1}+j_{1} v} \equiv l g^{i_{2}+j_{2} v}\text{ }\pmod{p},
$$
and hence
$$
\mathrm{ind}_{g}(s)+i_{1}+j_{1} v \equiv \mathrm{ind}_{g}(l)+i_{2}+j_{2} v \text{ }\pmod{p-1}.
$$
Reducing this to the residue modulo $v=2^{m-1}k$, we get
$$
\mathrm{ind}_{g}(s)+i_{1} \equiv \mathrm{ind}_{g}(l)+i_{2} \text{ }\pmod{v}.
$$
Since $g\equiv 1\pmod{2}$, then $s\equiv l\pmod{2}$ by (\ref{eqg}). Hence $s,l\in T_i\cup(-T_i)$. However, the two values $\mathrm{ind}_{g}(s) \pmod v$ and $\mathrm{ind}_{g}(l) \pmod v$  always belong to $\{\mathrm{ind}_g(x)\pmod{v}:\text{ }x\in T_i\}$ even when $s\in -T_i$ or $l\in -T_i$, due to the fact that $\mathrm{ind}_g(-1)\equiv \frac{p-1}{2}\pmod v\equiv 0\pmod v$. Then by the definition of $A$, we have $i_1=i_2$ as well as $s=l$ or $s=-l$.
\par
If $s=l$, then $j_1=j_2$.
\par
If $s=-l$, then $\frac{p-1}{2}+j_1 v\equiv j_2 v\pmod{p-1}$. Hence $\frac{p-1}{2}\mid v(j_1-j_2)$, that is $n\mid (j_1-j_2)$, which implies  $j_1=j_2$.
\end{pf}
\begin{rmk} It is not easy to generalize the construction in Theorem~\ref{dl7} to quasi-perfect $B[-k,k](tp)$ sets with $t>2$. To see this, let $k$ be a multiple of $t$, and we partition $[1,k]$ into $t$ residue classes modulo $t$. By the same arguments, we can deduce that $s\equiv l\pmod{t}$.  Then $s,l\in T_i\cup(-T_{t-i})$, from which we can not obtain the key conditions that $\mathrm{ind}_{g}(s) \pmod v$ and $\mathrm{ind}_{g}(l) \pmod v$  always belong to $\{\mathrm{ind}_g(x)\pmod{v}:\text{ }x\in T_i\}$.
\end{rmk}

\begin{example}
We give an example to compare the construction from Theorem \ref{dl7} and that from \cite[Theorem 5]{wz1}.
  Let $p=13729$, $k=8$, $m=1$. Then $g=23$ is a primitive root modulo $p$. We also have
  $$
  \begin{array}{llll}
  \mathrm{ind}_g(-8)=6654, & \mathrm{ind}_g(-7)=11084, & \mathrm{ind}_g(-6)=6376, & \mathrm{ind}_g(-5)=9594,\\
  \mathrm{ind}_g(-4)=11300, & \mathrm{ind}_g(-3)=11022, & \mathrm{ind}_g(-2)=2218, & \mathrm{ind}_g(-1)=6864,\\
  \mathrm{ind}_g(1)=0, & \mathrm{ind}_g(2)=9082, & \mathrm{ind}_g(3)=4158, & \mathrm{ind}_g(4)=4436,\\
  \mathrm{ind}_g(5)=2730, & \mathrm{ind}_g(6)=13240, & \mathrm{ind}_g(7)=4220, & \mathrm{ind}_g(8)=13518.
  \end{array}
  $$
It is easy to see that

\begin{align*}
   & \{\mathrm{ind}_g(i)\pmod{8}:\text{ }i=1,3,5,7\} \\
  = & \{\mathrm{ind}_g(i)\pmod{8}:\text{ }i=2,4,6,8\}=\{0,2,4,6\}.
\end{align*}
Then by Theorem \ref{dl7}, $\{23^{i+8j}\pmod{27458}:\text{ }i\in[0,1],j\in[0,857]\}$ is a quasi-perfect $B[-8,8](27458)$ set.
\par
Applying \cite[Theorem 5]{wz1} with $t=2$ and $\theta = \gcd \{ \mathrm{ind}_g( k ) \mid k \in [-8, 8 ]^*\} $, we get \[\left\{\frac{\mathrm{ind}_g(i)}{2}\pmod{8}:\text{ }i=\pm 1,\pm 3,\pm 5,\pm 7\right\}=\{0,5,6,7\}\] and
\[\left\{\frac{\mathrm{ind}_g(i)}{2}\pmod{8}:\text{ }i=\pm 2,\pm 4,\pm 6,\pm 8\right\}=\{2,4,5,7\}.\]However, both sets  have size $4\neq \frac{k_1+k_2}{t}=8$, hence we cannot get a quasi-perfect $B[-8,8](27458)$ set from  \cite[Theorem 5]{wz1}.
\end{example}
\begin{table}
  \centering
  \caption{Examples of quasi-perfect $B[-k,k](2p)$ sets from Theorem \ref{dl7}}
  \begin{tabular}{|c|c|c|}
    \hline
    $k$ & $m$ & $p$ \\
    \hline
    4 & 1 & 97, 241, 409, 457, 1009, 1129, 1489, 1873, 2017, 2161\\
    \hline
    4 & 2 & 577, 1201, 4801, 5233, 7393, 10513, 14401, 14449, 14593 \\
    \hline
    4 & 3 & 13441, 49633, 122497, 136993, 147457, 149377 \\
    \hline
    8 & 1 & 12721, 13729, 33889, 65809 \\
    \hline
  \end{tabular}
\end{table}

\subsection{Quasi-perfect $B[-(k-1),k](m)$ sets}
\begin{thm}\label{dl8}
  Let $k>0$ be an integer, and $p$ be a prime such that $k<p<\frac{4k-1}{3}$. Then
  $$
  B=\{k+1\}\cup\{1+(2k+2)i:\text{ }i\in[0,p-1]\}
  $$
  is a quasi-perfect $B[-(k-1),k](p(2k+2))$ set.
\end{thm}
\begin{pf}
  Suppose $r(k+1) \equiv s(k+1)\pmod{p(2 k+2)}$, where $r, s \in[-(k-1), k]^{*}$. Then $r \equiv s\pmod{2 p}$,  and so $r=s$.
\par
Suppose $r(k+1) \equiv s(1+(2 k+2) i)\pmod{p(2 k+2)}$, where $r, s \in[-(k-1), k]^{*}$ and $i \in[0, p-1]$. Then $s \equiv 0\pmod{k+1}$, which is a contradiction.
\par
Suppose $r(1+(2 k+2) i)\equiv s(1+(2 k+2) j)\pmod{p(2 k+2)}$, where $r, s \in[-(k-1), k]^{*}$ and $i,j \in[0, p-1]$. Then $r \equiv s\pmod{2 k+2}$, and so $r=s $. This implies $r(2 k+2) i\equiv r(2 k+2) j\pmod{p(2 k+2)}$, and so $r i \equiv r j\pmod{p}$. Note that $p>k$ is a prime, we have $\gcd(r, p)=1 $. Then $i \equiv j\pmod{p}$, and so $i=j$.
\par
Combining all pieces, we see that $B$ is a $B[-(k-1),k](p(2k+2))$ set of size $p+1=\left\lfloor\frac{p(2k+2)-1}{2k-1}\right\rfloor$.
\end{pf}
\begin{example}
\begin{enumerate}
  \item Let $k=6$ and $m=7$. By Theorem \ref{dl8}, the set
  $$\{1,7,15,29,43,57,71,85\}$$
  is a quasi-perfect $B[-5,6](98)$ set.
  \item Let $k=9$ and $m=11$. By Theorem \ref{dl8}, the set
  $$\{1,10,21,41,61,81,101,121,141,161,181,201\}$$
  is a quasi-perfect $B[-8,9](220)$ set.
\end{enumerate}
\end{example}

\section{Splitter Sets and Cayley Graphs}\label{sec5}
In this section, we give a connection between splitter sets and Cayley graphs. All the terminologies relevant to graph theory can be found in~\cite{bk5,bk6}. For the convenience of readers, we introduce some of them briefly.
\par
Suppose $H$ is a finite abelian group. Let $S$ be a subset of $H$ such that the identity $e\notin S$, and $s\in S$ implies that $s^{-1}\in S$. A \textit{Cayley graph} defined by $H$ and $S$ is an undirected graph $G=(V(G),E(G))$ with vertex set $V=H$ and edge set $E(G)$, such that $\{x,y\}\in E(G)$ if and only if $xy^{-1}\in S$. We denote it by $G=Cay(H,S)$. This kind of graph has been widely studied in the literature, such as \cite{wz26,wz27,wz28}.
\par
Given a graph $G=(V(G),E(G))$, a subset $I\subseteq V(G)$ is an \emph{independent} set if for any two distinct elements $x,y\in I$, $\{x,y\}\notin E(G)$. The maximum size of an independent set is called the \emph{independence number}, denoted as $\alpha(G)$. We say $G$ is \emph{$d$-regular}, if for each $x\in V(G)$, there exist exactly $d$ vertices $y\in V(G)$ such that $\{x,y\}\in E(G)$. We say a sequence of pairwise-distinct vertices $P=x_0x_1\cdots x_{n-1}x_n$ ($n\ge 1$) is a \emph{path} connecting $x_0$ and $x_n$, if $\{x_i,x_{i+1}\}\in E(G)$ for any $i=0,\ldots,n-1$. If for any distinct $x,y\in V(G)$, there is a path connecting $x$ and $y$, we say $G$ is \emph{connected}. A maximal connected subgraph of $G$ is called a \emph{connected component}. A $2$-regular connected graph is called a \emph{cycle}. We say two graphs $G_1=(V_1,E_1)$ and $G_1=(V_2,E_2)$ are \emph{isomorphic}, if there exists a bijection $f:V_1\longrightarrow V_2$ such that $\{x,y\}\in E_1$ if and only if $\{f(x),f(y)\}\in E_2$.
\par
In the rest of this section, we let $ k_2\geq k_1\geq 0$ be integers and $M=[-k_1,k_2]^{*}$. Since perfect $B[0,1](p)$ sets and perfect $B[-1,1](p)$ sets are trivial, and maximal $B[-k_1,2](q)$ sets have been completely determined for any $q$ in~\cite{wz2,wz7,wz17}, we assume $k_2\ge 3$ in this section. For any prime $p>k_1+k_2$, $M$ can be seen as a subset of $\mathbb{Z}_p^{*}$. Let $S=\{xy^{-1}:\text{ }x,y\in M\text{ and }x\ne y\}$, $G=Cay(\mathbb{Z}_p^{*},S)$ and $G^{\prime}=Cay(\langle M\rangle,S)$. Note that $\langle S\rangle=\langle M\rangle$, so $G^{\prime}$ is a connected component of $G$~\cite{wz28} and each connected component of $G$ is isomorphic to $G^{\prime}$.
\par
First, we have the following observation.
\begin{prop}\label{ind}
A subset $B\subset\mathbb{Z}_p^{*}$ is a $B[-k_1,k_2](p)$ set if and only if $B$ is an independent set in $G$.
\end{prop}
\begin{pf}
  First, suppose $B$ is a $B[-k_1,k_2](p)$ set. If there exist two different $b_1,b_2\in B$ such that $\{b_1,b_2\}\in E(G)$, then there exist two different $x,y\in M$ such that $b_1b_2^{-1}=xy^{-1}$, i.e. $xb_2=yb_1$. Since $B$ is a $B[-k_1,k_2](p)$ set, we have $x=y$ and $b_1=b_2$, which is a contradiction. So $B$ is an independent set in $G$.
  \par
  On the other hand, suppose $B$ is an independent set in $G$. If there exist $b_1,b_2\in B$ and $x,y\in M$ such that $xb_1=yb_2$, then $b_1b_2^{-1}=yx^{-1}$. If $b_1\neq b_2$, then by the definition of $G$, $\{b_1,b_2\}\in E(G)$, which contradicts the assumption that $B$ is an independent set in $G$. So $b_1=b_2$ and $x=y$. Thus $B$ is a $B[-k_1,k_2](p)$ set.
\end{pf}
By Proposition~\ref{ind},  a $B[-k_1,k_2](p)$ set of maximum size is equivalent to  a maximum independent set in the graph $G$. The next lemma is a corollary of Brooks' theorem \cite{wz6}. It can be used to give a nontrivial lower bound on the size of a maximum $B[-k_1,k_2](p)$ set for any prime $p>k_1+k_2+1$ and any $0\le k_1\le k_2$ with $k_2\ge 3$. To the best of our knowledge, there was no general lower bound before. Recall that a complete graph is a graph $\Gamma$ in which $\{x,y\}\in E(\Gamma)$ for each pair of distinct $x,y\in V(\Gamma)$, and an odd cycle is a cycle with odd vertices.
\begin{lem}
  Let $\Gamma$ be a $d$-regular graph. If each connected component of $\Gamma$ is not a complete graph or an odd cycle, then
  $$
  \alpha(\Gamma)\ge\frac{|V(\Gamma)|}{d}.
  $$
Otherwise,
$$
  \alpha(\Gamma)\ge\frac{|V(\Gamma)|}{d+1}.
$$
\end{lem}
From the definition, we can easily check that $G$ and $G^{\prime}$ are both $|S|$-regular. If $p>k_1+k_2+2$ and $k_2\ge 3$, then $|S|>2$ and therefore $G^{\prime}$ is not an odd cycle. Since $G^{\prime}$ is $|S|$-regular, we have $|\langle M\rangle|\ge |S|+1$. Furthermore, if $|\langle M\rangle|\ge |S|+2$, $G^{\prime}$  cannot be a complete graph. Thus, we have the following corollary.
\begin{cor}\label{corollary2}
  Let $B$ be a $B[-k_1,k_2](p)$ set of maximum size. If $p>k_1+k_2+1$ and $k_2\ge 3$, then
  $$
  |B|\ge\left\lceil\frac{p-1}{|S|+1}\right\rceil.
  $$
  Further, if $|\langle M\rangle|\ge |S|+2$ and $p>k_1+k_2+2$, then
  $$
  |B|\ge\left\lceil\frac{p-1}{|S|}\right\rceil.
  $$
\end{cor}
There is another advantage by connecting splitter sets with Cayley graphs: we can use some mathematical softwares such as Maple to get a maximum independent set of graphs (and thus a splitter set of maximum size).
\begin{example}
Take $k_1=0,k_2=3$, we compute some values listed in Table~\Rnum{4} below. The third row is a lower bound from Corollary~\ref{corollary2}.
The fourth row is computed via the command \emph{IndependenceNumber} in Maple and the last row is a maximum independent set (i.e. a splitter set of maximum size) computed via the command \emph{MaximumIndependentSet} in Maple.
\begin{table}[htbp]\label{s4}
\centering
\caption{The case $k_1=0$, $k_2=3$}
\begin{tabular}{|c|c|c|c|c|c|c|c|c|c|}
  \hline
   $p$&7&11&13&17&19&23&29&31&37\\
   \hline
   $|S|$&4&6 &6 &6 &6 &6 &6 &6 &6 \\
   \hline
   Corollary~\ref{corollary2}&2&2&2&3&3&4&5&5&6\\
   \hline
   $\alpha(G)$&2&2&3&4 &5 &5 &8 &8 &12 \\
   \hline
   \tabincell{c}{a maximum\\ independent set\\ (maximum \\splitter set)}&\{1,6\}&\{1,5\}&\{1,4,11\}&\{1,4,13,16\}&\{1,6,8,14,15\}&\{1,4,5,6,7\}&\tabincell{c}{\{1,5,6,7,8,\\~~11,19,26\}}&\tabincell{c}{$\{1,4,9,10,14$,\\$23,25,26$\}}&\tabincell{c}{$\{1,6,8,10,11$,\\$~~14,23,26,27$,\\$29,31,36$\}}\\
   \hline
\end{tabular}
\end{table}
\end{example}

\section{Conclusion}\label{sec6}
In this paper, we  consider the existence  of  splitter sets. We give some necessary and sufficient conditions for
the existence of a nonsingular perfect $B[-k_1,k_2](p)$ set, where $(k_{1},k_{2})\in\{(0,4),(2,4),(4,4)\}$. For easy reference, we summarize the equivalent conditions obtained in this paper and related known results in Table \ref{table:summary}, where $p$ is a prime, $g$ is a primitive root modulo $p$ and $\mu=\gcd\{\mathrm{ind}_g(j):\text{ }j\in [-1,k]^{*}\}$ (for $B[-k,k](p)$ sets), or $\mu=\gcd\{\mathrm{ind}_g(j):\text{ }j\in \{2,\ldots,k,p-1\}\}$ (for $B[0,k](p)$ sets),  or $\mu=\gcd\{\mathrm{ind}_g(j):\text{ }j\in [-1,k_2]^{*}\}$ (for $B[-k_1,k_2](p)$ sets).
\begin{table}[tbhp]
\begin{center}
\caption{Existence of nonsingular perfect splitter sets}
\label{table:summary}
\begin{tabular}{|c|l|l|}
\hline
Nonsingular perfect splitter sets &
Necessary and sufficient conditions & Remarks
\\ \hline
$B[-k,k](p)$, where $k$ is an odd prime & $p\equiv 1\pmod{2\mu k}$ and $\left|\left\{\frac{\mathrm{ind}_{g}(j)}{\mu}\pmod{k} : j \in[1, k]\right\}\right|=k$ & Theorem 3.2 of \cite{wz23}\\
\hline
$B[0,k](p)$, where $k$ is an odd prime & $p\equiv 1\pmod{\mu k}$ and $\left|\left\{\frac{\mathrm{ind}_{g}(j)}{\mu}\pmod{k} : j \in[1, k]\right\}\right|=k$ & Theorem 3.3 of \cite{wz23}\\
\hline
$B[-k_1,k_2](p)$, $\gcd(\frac{p-1}{k_1+k_2},k_1+k_2)=1$ & \tabincell{l}{$p\equiv 1\pmod{\mu(k_1+k_2)}$ and \\ $\left|\left\{\frac{\mathrm{ind}_{g}(j)}{\mu}\pmod{k} : j \in[-k_1, k_2]^{*}\right\}\right|=k_1+k_2$} & Theorem 3.5 of \cite{wz23}\\
\hline
$B[0,2](p)$ & $p\equiv 1\pmod{2}$ and $\mathrm{ord}_p(2)$ is even & Theorem 2 of \cite{wz5}\\
\hline
$B[-2,2](p)$ & $p\equiv 1\pmod{4}$ and $v_2(\mathrm{ord}_p(2))\ge 2$ & Corollary 3 of \cite{wz17}\\
\hline
\multirow{2}{*}{$B[-1,3](p)$}
 & $p\equiv 5\pmod{8}$, $6$ is a quartic residue modulo $p$ & Theorem 4.4 of \cite{wz4}\\ \cline{2-3}
 &$p\equiv 1\pmod{8}$, $\mathrm{ord}_p(-\frac{3}{2})$ is odd and $4\mid\mathrm{ord}_p(2)$ & Theorem 4.5 of \cite{wz4}\\
\hline
$B[-2,4](p)$ & $p\equiv 1\pmod{6}$, $\mathrm{ord}_p(-\frac{3}{4})$ is odd and $2\notin\langle 6,8\rangle$ & Theorem~\ref{thm0}\\
\hline
$B[-4,4](p)$ & $p\equiv 1\pmod{8}$ and $\pm 4\notin\langle 6,16\rangle$& Theorem~\ref{thm1}\\
\hline
$B[0,4](p)$  & $p\equiv 1\pmod{4}$ and $4\notin\langle 6,16\rangle$ & Theorem~\ref{thm2}\\
\hline
\end{tabular}
\end{center}
\end{table}

We also present four new constructions of quasi-perfect splitter sets. Finally, we give a general lower bound on the maximum size of a $B[-k_1,k_2](p)$ set for any prime $p>k_1+k_2+1$ and any $k_2\geq k_1\geq 0$, by connecting splitter sets with independent sets of Cayley graphs.
\par
For future work, we suggest the following questions.
\begin{enumerate}
  \item Prove the nonexistence conjectures for purely {\it singular} perfect splitter sets proposed in \cite{wz22,wz23}.
  \item Determine the maximum size of $B[-k_1, k_2](n)$ sets. This problem has been completely solved for $0\le k_{1}\le k_{2}\le2$ \cite{wz5,wz17,wz2}.
  \item Give a characterization of nonsingular perfect $B[-k_1, k_2](p)$ sets. In \cite{wz23}, the authors proved that there does not exist a nonsingular perfect $B[-k_1, k_2](p)$ set when $1\le k_{1}< k_{2}$ and $k_{1}+k_{2}$ is odd. The other results are listed in Table~\ref{table:summary}. In this paper, we completely determine the condition for the existence of a nonsingular perfect $B[-k_1, k_2](p)$ set, where  $(k_{1},k_{2})\in\{(0,4),(2,4),(4,4)\}$. The next case is $(k_{1},k_{2})=(1,5)$.
  \item Give more constructions of perfect or quasi-perfect splitter sets. In \cite[Table \Rnum{5}]{wz17}, the authors listed $B[-3,3](q)$ sets of maximum size for all $q \leq 70$. Among these, there are eight nontrivial perfect or quasi-perfect $B[-3,3](q)$ sets, six of which are examples obtained from general theorems in the same paper. In this paper, we give a construction of quasi-perfect $B[-k, k](p(2 k+2))$ sets in Theorem~\ref{dl6}, where $p\in[k+1,2k-1]$ is a prime. This gives a quasi-perfect $B[-3,3](40)$ set $\{1,4,9,17,25,33\}$ as in Example~\ref{ex1}, which is different from the one $\{1,4,5,7,9,17\}$ given in \cite[Table \Rnum{5}]{wz17}. There is still one more quasi-perfect splitter set given in \cite[Table \Rnum{5}]{wz17}, that is, $B[-3,3](18)=\{1,4\}$. We wonder whether this example could be generalized to an infinite family.
  \item Find more constructions of  splitter sets of maximum size. One may try to generalize the splitter sets listed in Table \Rnum{5} of \cite{wz17}.
\end{enumerate}
\section*{Acknowledgement}
The authors express their gratitude to the anonymous reviewers for their detailed and constructive comments which are very helpful to the improvement of the presentation of this paper. In particular, we thank one of the reviewers for introducing us the reference \cite{M95}. The authors would also like to thank Prof. Moshe Schwartz, the associate editor, for his excellent editorial job.

\bibliographystyle{IEEEtranS}
\bibliography{New_Results_on_Splitter_on_Sets}

% Generated by IEEEtranS.bst, version: 1.13 (2008/09/30)
\begin{thebibliography}{10}
\providecommand{\url}[1]{#1}
\csname url@samestyle\endcsname
\providecommand{\newblock}{\relax}
\providecommand{\bibinfo}[2]{#2}
\providecommand{\BIBentrySTDinterwordspacing}{\spaceskip=0pt\relax}
\providecommand{\BIBentryALTinterwordstretchfactor}{4}
\providecommand{\BIBentryALTinterwordspacing}{\spaceskip=\fontdimen2\font plus
\BIBentryALTinterwordstretchfactor\fontdimen3\font minus
  \fontdimen4\font\relax}
\providecommand{\BIBforeignlanguage}[2]{{%
\expandafter\ifx\csname l@#1\endcsname\relax
\typeout{** WARNING: IEEEtranS.bst: No hyphenation pattern has been}%
\typeout{** loaded for the language `#1'. Using the pattern for}%
\typeout{** the default language instead.}%
\else
\language=\csname l@#1\endcsname
\fi
#2}}
\providecommand{\BIBdecl}{\relax}
\BIBdecl

\bibitem{wz27}
L.~Babai, ``Spectra of {C}ayley graphs,'' \emph{Journal of Combinatorial
  Theory. Series B}, vol.~27, no.~2, pp. 180--189, 1979.

\bibitem{wz21}
S.~R. Blackburn and J.~F. McKee, ``Constructing $k$-radius sequences,''
  \emph{Mathematics of Computation}, vol.~81, no. 280, pp. 2439--2459, 2012.

\bibitem{bk6}
A.~{Bondy} and M.~{Murty}, \emph{Graph Theory}, 1st~ed., ser. Graduate Texts in
  Mathematics.\hskip 1em plus 0.5em minus 0.4em\relax Springer-Verlag London,
  2008, vol. 244.

\bibitem{wz15}
S.~{Buzaglo} and T.~{Etzion}, ``Tilings with $n$-dimensional chairs and their
  applications to asymmetric codes,'' \emph{IEEE Transactions on Information
  Theory}, vol.~59, no.~3, pp. 1573--1582, 2013.

\bibitem{wz8}
Y.~{Cassuto}, M.~{Schwartz}, V.~{Bohossian}, and J.~{Bruck}, ``Codes for
  asymmetric limited-magnitude errors with application to multilevel flash
  memories,'' \emph{IEEE Transactions on Information Theory}, vol.~56, no.~4,
  pp. 1582--1595, 2010.

\bibitem{web1}
W.~contributors, ``Bunyakovsky conjecture,''
  \url{https://en.wikipedia.org/w/index.php?title=Bunyakovsky_conjecture&oldid=898182463},
  date of last revision: 21 May 2019 22:43 UTC.

\bibitem{wz16}
N.~{Elarief} and B.~{Bose}, ``Optimal, systematic, $q$-ary codes correcting all
  asymmetric and symmetric errors of limited magnitude,'' \emph{IEEE
  Transactions on Information Theory}, vol.~56, no.~3, pp. 979--983, 2010.

\bibitem{bk5}
C.~{Godsil} and G.~{Royle}, \emph{Algebraic Graph Theory}, 1st~ed., ser.
  Graduate Texts in Mathematics.\hskip 1em plus 0.5em minus 0.4em\relax
  Springer-Verlag New York, 2001, vol. 207.

\bibitem{wz9}
D.~Hickerson and S.~Stein, ``Abelian groups and packing by semicrosses,''
  \emph{Pacific Journal of Mathematics}, vol. 122, no.~1, pp. 95--109, 1986.

\bibitem{wz26}
W.~Imrich, ``On the connectivity of {C}ayley graphs,'' \emph{Journal of
  Combinatorial Theory. Series B}, vol.~26, no.~3, pp. 323--326, 1979.

\bibitem{bk2}
K.~Ireland and M.~Rosen, \emph{A Classical Introduction to Modern Number
  Theory}, 2nd~ed., ser. Graduate Texts in Mathematics.\hskip 1em plus 0.5em
  minus 0.4em\relax Springer-Verlag New York, 1990, vol.~84.

\bibitem{wz7}
T.~{Kl\o ve}, B.~{Bose}, and N.~{Elarief}, ``Systematic, single limited
  magnitude error correcting codes for flash memories,'' \emph{IEEE
  Transactions on Information Theory}, vol.~57, no.~7, pp. 4477--4487, 2011.

\bibitem{wz5}
T.~{Kl\o ve}, J.~{Luo}, I.~{Naydenova}, and S.~{Yari}, ``Some codes correcting
  asymmetric errors of limited magnitude,'' \emph{IEEE Transactions on
  Information Theory}, vol.~57, no.~11, pp. 7459--7472, 2011.

\bibitem{wz17}
T.~{Klove}, J.~{Luo}, and S.~{Yari}, ``Codes correcting single errors of
  limited magnitude,'' \emph{IEEE Transactions on Information Theory}, vol.~58,
  no.~4, pp. 2206--2219, 2012.

\bibitem{wz6}
L.~Lov\'{a}sz, ``Three short proofs in graph theory,'' \emph{Journal of
  Combinatorial Theory, Series B}, vol.~19, no.~3, pp. 269--271, 1975.

\bibitem{wz18}
S.~{Martirossian}, ``Single-error correcting close packed and perfect codes,''
  in \emph{Proceedings of 1st INTAS International Seminar on Coding Theory and
  Combinatorics}, 1996, pp. 90--115.

\bibitem{M95}
A.~Munemasa, ``On perfect {$t$}-shift codes in abelian groups,'' \emph{Des.
  Codes Cryptogr.}, vol.~5, no.~3, pp. 253--259, 1995.

\bibitem{wz19}
M.~{Schwartz}, ``Quasi-cross lattice tilings with applications to flash
  memory,'' \emph{IEEE Transactions on Information Theory}, vol.~58, no.~4, pp.
  2397--2405, 2012.

\bibitem{wz20}
------, ``On the non-existence of lattice tilings by quasi-crosses,''
  \emph{European Journal of Combinatorics}, vol.~36, pp. 130--142, 2014.

\bibitem{wz28}
M.~Shahzamanian, M.~Shirmohammadi, and B.~Davvaz, ``Roughness in cayley
  graphs,'' \emph{Information Sciences}, vol. 180, no.~17, pp. 3362--3372,
  2010.

\bibitem{wz10}
S.~{Stein}, ``Factoring by subsets,'' \emph{Pacific Journal of Mathematics},
  vol.~22, no.~3, pp. 523--541, 1967.

\bibitem{wz11}
------, ``Packings of $\mathbb{R}^{n}$ by certain error spheres,'' \emph{IEEE
  Transactions on Information Theory}, vol.~30, no.~2, pp. 356--363, 1984.

\bibitem{bk4}
S.~Stein and S.~Szab{\'o}, \emph{Algebra and Tiling: Homomorphisms in the
  Service of Geometry}, ser. Carus Mathematical Monographs.\hskip 1em plus
  0.5em minus 0.4em\relax Mathematical Association of America, 1994, vol.~25.

\bibitem{bk1}
S.~Szab\'{o} and A.~Sands, \emph{Factoring Groups into Subsets}, ser. Lecture
  Notes in Pure and Applied Mathematics.\hskip 1em plus 0.5em minus 0.4em\relax
  Boca Raton, FL, USA: CRC Press, 2009, vol. 257.

\bibitem{wz12}
S.~Szab{\'o}, ``Some problems on splittings of groups,'' \emph{Aequationes
  Mathematicae}, vol.~30, no.~1, pp. 70--79, Dec 1986.

\bibitem{wz13}
------, ``Some problems on splittings of groups
  \uppercase\expandafter{\romannumeral2},'' \emph{Proceedings of the American
  Mathematical Society}, vol. 101, no.~4, pp. 585--591, 1987.

\bibitem{wz14}
U.~{Tamm}, ``Splittings of cyclic groups and perfect shift codes,'' \emph{IEEE
  Transactions on Information Theory}, vol.~44, no.~5, pp. 2003--2009, 1998.

\bibitem{wz22}
A.~J. Woldar, ``A reduction theorem on purely singular splittings of cyclic
  groups,'' \emph{Proceedings of the American Mathematical Society}, vol. 123,
  no.~10, pp. 2955--2959, 1995.

\bibitem{wz24}
D.~Xie and J.~Luo, ``Asymmetric single magnitude four error correcting codes,''
  \emph{CoRR}, vol. abs/1903.01148, 2019.

\bibitem{wz25}
------, ``Correcting codes for asymmetric single magnitude four error,''
  \emph{CoRR}, vol. abs/1905.02570, 2019.

\bibitem{wz2}
S.~{Yari}, T.~{Kl\o ve}, and B.~{Bose}, ``Some codes correcting unbalanced
  errors of limited magnitude for flash memories,'' \emph{IEEE Transactions on
  Information Theory}, vol.~59, no.~11, pp. 7278--7287, 2013.

\bibitem{wz4}
P.~Yuan and K.~Zhao, ``On the existence of perfect splitter sets,''
  \emph{CoRR}, vol. abs/1903.00118, 2019.

\bibitem{wz1}
T.~{Zhang} and G.~{Ge}, ``New results on codes correcting single error of
  limited magnitude for flash memory,'' \emph{IEEE Transactions on Information
  Theory}, vol.~62, no.~8, pp. 4494--4500, 2016.

\bibitem{wz23}
------, ``On the nonexistence of perfect splitter sets,'' \emph{IEEE
  Transactions on Information Theory}, vol.~64, no.~10, pp. 6561--6566, 2018.

\bibitem{wz3}
T.~{Zhang}, X.~{Zhang}, and G.~{Ge}, ``Splitter sets and $k$-radius
  sequences,'' \emph{IEEE Transactions on Information Theory}, vol.~63, no.~12,
  pp. 7633--7645, 2017.

\end{thebibliography}
\end{document}